\documentclass[prd,10pt,twocolumn,showpacs,nofootinbib, aps,superscriptaddress,preprintnumbers]{revtex4-1}

\usepackage{bm}
\usepackage{graphicx}
\usepackage{multirow}
\usepackage{amsmath}
\usepackage{mathrsfs}
\usepackage{amssymb}
\usepackage{hyperref}\usepackage{yfonts}
\usepackage{color}
\usepackage{xspace}
\usepackage{verbatim}
\usepackage{mathtools}
\usepackage{booktabs}
\usepackage{tabularx}
\usepackage[utf8x]{inputenc} 
\usepackage[dvipsnames,table]{xcolor}
\definecolor{lightgray}{rgb}{0.9,0.9,0.9}	    
\definecolor{green}{rgb}{0,0.5,0}
\definecolor{red}{rgb}{1,0,0}
\definecolor{blue}{rgb}{0,0,0.5}

\def\araa{{ARA\&A}}

\newcommand{\qhat}{\hat{\mathbf{q}}}

\newcommand{\vmin}{v_{\rm min}}

\newcommand{\vesc}{v_\mathrm{esc}}
\newcommand{\GeVcm}{\textrm{ GeV cm}^{-3}}
\newcommand{\kms}{\textrm{ km s}^{-1}}

\newcommand{\dbd}[2]{\ifmmode \frac{\textrm{d}#1}{\textrm{d}#2}\else $\textrm{d}#1/\textrm{d}#2$\fi}
\newcommand{\pbp}[2]{\ifmmode \frac{\partial#1}{\partial#2}\else $\partial#1/\partial#2$\fi}
\newcommand{\erf}{\mathrm{erf}}
\newcommand{\erfi}{\mathrm{erfi}}
\newcommand{\ra}[1]{\renewcommand{\arraystretch}{#1}}

\newcommand{\Msol}{\, \textrm{M}_\odot}
\newcommand{\vbf}{\mathbf{v}}

\DeclareMathAlphabet{\mathpzc}{OT1}{pzc}{m}{it}

\newcommand{\fR}{f_{\rm R}}
\newcommand{\fS}{f_{\rm S}}
\newcommand{\rhoS}{\rho_{\rm S}}
\newcommand{\SHMpp}{SHM$^{++}$}

\begin{document}

\title{SHM$^{++}$: A Refinement of the Standard Halo Model for Dark
  Matter Searches}

\author{N. Wyn Evans} \email{nwe@ast.cam.ac.uk} \affiliation{Institute
  of Astronomy, Madingley Rd, Cambridge, CB3 0HA, United Kingdom}

\author{Ciaran A. J. O'Hare} \email{ciaran.aj.ohare@gmail.com}
\affiliation{Departamento de F\'isica Te\'orica, Universidad de
  Zaragoza, Pedro Cerbuna 12, E-50009, Zaragoza, Espa\~{n}a}

\author{Christopher McCabe} \email{christopher.mccabe@kcl.ac.uk}
\affiliation{Department of Physics, King's College London, Strand,
  London, WC2R 2LS, United Kingdom}
  
  \preprint{KCL-PH-TH-2018-49}

\date{\today}
\smallskip
\begin{abstract}
Predicting signals in experiments to directly detect dark matter (DM)
requires a form for the local DM velocity distribution.  Hitherto, the
standard halo model (SHM), in which velocities are isotropic and
follow a truncated Gaussian law, has performed this job.  New data,
however, suggest that a substantial fraction of our stellar halo lies
in a strongly radially anisotropic population, the `\emph{Gaia}
Sausage'.  Inspired by this recent discovery, we introduce an updated
DM halo model, the SHM$^{++}$, which includes a `Sausage' component,
thus better describing the known features of our galaxy.  The
SHM$^{++}$ is a simple analytic model with five parameters: the
circular speed, local escape speed and local DM density, which we
update to be consistent with the latest data, and two new parameters:
the anisotropy and the density of DM in the Sausage.  The impact of
the SHM$^{++}$ on signal models for WIMPs and axions is rather modest
since the multiple changes and updates have competing effects.  In
particular, this means that the older exclusion limits derived for
WIMPS are still reasonably accurate.  However, changes do occur for
directional detectors, which have sensitivity to the full
three-dimensional velocity distribution.
\end{abstract}

\maketitle

\section{Introduction}

Historically, analyses of direct searches for dark matter (DM) have
constructed signal models based upon the Gaussian distribution of
velocities found in the standard halo model (SHM). This is inspired by isothermal
spheres, which have asymptotically flat rotation curves. They are the
only exact, self-gravitating systems with Gaussian velocity
distributions~\cite{Chandra}. Of course, it has long been known that
the SHM is an
idealisation~\cite{Ev06,Vogelsberger:2008qb,Ze09,Ku10,Mao:2013nda},
but it provides an excellent trade-off between simplicity and
realism. The effects of non-Maxwellian speed
distributions~\cite{Kuhlen:2009vh,Ling:2009eh,Mao:2012hf},
triaxiality~\cite{Ca00}, velocity
anisotropy~\cite{Fairbairn:2008gz,Knirck:2018knd,MarchRussell:2008dy,Bozorgnia:2013pua,Fornasa:2013iaa},
streams~\cite{Savage:2006qr,Lee:2012pf,Purcell:2012sh,O'Hare:2014oxa,OHare:2017yze,Foster:2017hbq},
and other dark
substructures~\cite{Bruch:2008rx,Lisanti:2010qx,Lisanti:2011as,Billard:2012qu,Kavanagh:2016xfi}
have all received attention using simple elaborations of the
SHM. These studies were speculative and theoretically motivated given
that in the past there was sparse knowledge about the true DM velocity
distribution. While it is possible, and in some cases advisable, to
derive exclusion limits on DM particle physics without any
astrophysical assumptions, e.g. Refs.~\cite{Fox:2010bz, McCabe:2011sr,
  Frandsen:2011gi, Gondolo:2012rs, HerreroGarcia:2012fu,
  Frandsen:2013cna, Bozorgnia:2013hsa, Feldstein:2014gza,
  Fox:2014kua,Feldstein:2014ufa, Anderson:2015xaa,
  Gelmini:2016pei,Kahlhoefer:2016eds,Gondolo:2017jro,Gelmini:2017aqe,Ibarra:2017mzt,Fowlie:2017ufs},
this often comes at the cost of greater complexity and less
overall constraining power.

The arrival of the second data release from the {\it Gaia}
satellite has been transformational for our understanding of the
structure of the Galaxy~\cite{GaiaDR2}. The shape of the stellar halo,
the local DM density, the local circular speed, the escape velocity
and the history of accretion have all been the subject of sometimes
radical revision in the wake of the new and abundant
data~\cite{Lancaster:2018,We18,Be18}. Our understanding of the DM
halo has not been left unscathed by the \emph{Gaia} Revolution, and
the time is ripe to put forward a new standard halo model, SHM$^{++}$,
that represents our current knowledge, yet rivals the SHM in
simplicity and realism.

The most substantial change brought about by \emph{Gaia} data is that
the local stellar halo is now known to have two
components~\cite{Ca07,My18,Be18,Ma18}. The more metal-poor stars form
a weakly rotating structure that is almost spherical (with axis ratio
$q \approx 0.9$). This is likely the residue of many ancient
accretions from low mass dwarf galaxies in random directions so that
the net angular momentum of the accumulated material is almost
zero. The more metal-rich stars form a flattened ($q \approx 0.6$),
highly radially anisotropic structure. This is the ``\emph{Gaia}
Sausage''. It was created by the more recent accretion of a large
dwarf galaxy of mass $\approx 10^{10} - 10^{11} \Msol$ around 8 to 10 billion
years ago~\citep{Be18,He18,Kr18}, which will have been accompanied by
a corresponding avalanche of DM.

Given what we now know about the stellar halo, it is natural to expect
that the local DM halo also has a bimodal structure, made up from a
rounder, isotropic component with velocity distribution $\fR$ and a
radially anisotropic Sausage component $\fS$.  In
ref.~\citep{Necib:2018iwb}, the velocity distributions of these two
components were inferred from the velocities of stellar populations.
Here, we provide simple analytic velocity distributions that capture
the generic features of both components. The fraction of the local DM
in the Sausage $\eta$ is not well known, though we will argue that it
lies between 10\% and 30\%.  The velocity anisotropy of DM in the
Sausage~$\beta$ is also not known, but the stellar and globular
cluster populations associated with the Sausage are all extremely
eccentric and so must be the~DM.

In Section~\ref{sec:SHM}, we discuss the shortcomings of the SHM in
the light of recent advances in our knowledge of Galactic
structure. Section~\ref{sec:SHMpp} introduces the SHM$^{++}$, which
acknowledges explicitly the bimodal structure of the Galaxy's dark
halo.  We also take the opportunity to update the Galactic constants
in the SHM$^{++}$, as the familiar choices for the SHM represent the
state of knowledge that is now over a decade or more old. In
Section~\ref{sec:alternatives}, we discuss how our model compares with
other complementary strategies for determining the local velocity
distribution of DM. Then, Section~\ref{sec:dm} discusses the
implications for a range of WIMP and axion direct detection
experiments. We sum up in Section~\ref{sec:conc}.

\section{The SHM: A Critical Discussion}\label{sec:SHM}
At large radii the rotation curve of the Milky Way is flat to a good approximation~\citep{So09}. The family of isothermal spheres (of which
the most familiar example is the singular isothermal sphere) provide
the simplest spherical models with asymptotically flat rotation
curves~\cite{BT}. These models all have Gaussian
velocity distributions.

The SHM was introduced into astroparticle physics over thirty years
ago~\citep{Dr86}. It models a smooth round dark halo. The
velocity distribution for DM is a Gaussian in the Galactic frame, namely
\begin{align}\label{eq:shm}
\fR(\vbf) = \frac{1}{(2\pi
  \sigma_v^2)^{3/2}N_\mathrm{R,esc}} \, \exp \left( - \frac{|\vbf|^2}{2\sigma_v^2}\right) \,
\nonumber\\ \times \Theta (\vesc - |\vbf|)\,,
\end{align}
where~$\sigma_v$ is the isotropic velocity dispersion of the DM and
$v_0 = \sqrt{2} \sigma_v$ is the value of the asymptotically flat
rotation curve.  The isothermal spheres all have infinite extent,
whereas Galaxy halos are finite. This is achieved in the SHM by
truncating the velocity distribution at the escape speed~$\vesc$,
using the Heaviside function~$\Theta$. The constant $N_\mathrm{R,
  esc}$ is used to renormalize the velocity distribution after
truncation,
\begin{equation}\label{eq:norm}
N_\mathrm{R,esc} = \erf \left( \frac{\vesc}{\sqrt{2}\sigma_v}\right) -
\sqrt{\frac{2}{\pi}} \frac{\vesc}{\sigma_v} \exp \left(
-\frac{\vesc^2}{2\sigma_v^2} \right)\,.
\end{equation}

Hence to describe the velocity distribution of DM in the galactic
frame under the SHM we only need to prescribe two parameters, $v_0$
and $\vesc$. The value of $v_0$ is usually taken as equivalent to the
velocity of the Local Standard of Rest (or the circular velocity at
the Solar position). The assumed value of $\vesc$ has also typically
been inspired by various astronomical determinations.  The standard
values for these quantities in the SHM are listed in
Table~\ref{tab:astrobenchmarks}. These values are, however,
now somewhat out of date having undergone significant revision in
recent years. One motivation for updating the SHM is to
incorporate the more recent values for these parameters.

\begin{table}[t!]
\ra{1.3}
\begin{tabular}{@{}llll@{}}
\hline\hline
\multirow{5}{*}{\bf SHM} & Local DM density &$\rho_0$ 		& $0.3 \, \mathrm{GeV} \, \mathrm{cm}^{-3}$\\
			& Circular rotation speed &$v_0$ 	& $220 \kms$\\
			& Escape speed &$\vesc$ & $544 \kms$\\
			& Velocity distribution & $\fR(\mathbf{v})$ & Eq.~\eqref{eq:shm} \\
\hline
\multirow{6}{*}{{\bf SHM}$^{++}$} & Local DM density &$\rho_0$ & $0.55\pm 0.17\GeVcm$\\
			& Circular rotation speed & $v_0$ 	& $233\pm 3 \kms$\\
			& Escape speed &$\vesc$ & $528^{+24}_{-25} \kms$\\
			& Sausage anisotropy & $\beta$ & $0.9\pm 0.05$ \\
				& Sausage fraction &$\eta$ & $0.2 \pm 0.1$\\
				& Velocity distribution & $f(\mathbf{v})$ & Eq.~\eqref{eq:shmpp} \\
\hline\hline
\end{tabular}
\caption{The astrophysical parameters and functions defining the SHM
  and the SHM$^{++}$. We include a recommendation for the uncertainty
  on each parameter for analyses that incorporate astrophysical
  uncertainties. While the uncertainties associated with $\rho_0$,
  $v_0$ and $\vesc$ are based on direct measurements, the
  uncertainties associated with $\beta$ and $\eta$ are less
  certain. We refer the reader to the discussion in Section IIIA and
  IIIB respectively for more details.
}
\label{tab:astrobenchmarks}
\end{table}

The SHM has some successful features that we want to maintain.
Current theories of galaxy formation in the cold dark
matter paradigm envisage the build-up of DM halos through
accretion and merger. 
In the inner halo (where the Sun is located), the distribution of DM
particles extrapolated via sub-grid methods in high resolution
dissipationless simulations like Aquarius is rather
smooth~\citep{Vo11}, so a smooth velocity distribution
is a good assumption.
Furthermore, recent
hydrodynamic simulations~\cite{Bozorgnia:2016ogo,Kelso:2016qqj,Sloane:2016kyi,Lentz:2017aay} have recovered speed distributions for DM
that are better approximated by Maxwellian-distributions than their earlier N-body
counterparts~\cite{Hansen:2005yj,Vogelsberger:2008qb,Kuhlen:2009vh,Ling:2009eh,Mao:2012hf,Mao:2013nda}. 
In this light, the assumption in the SHM of a Gaussian velocity distribution
 is surprisingly accurate.

There is, however, a significant shortcoming to the SHM.
\emph{Gaia} data has provided significant new information 
about the stellar and dark halo of our own Galaxy.
The halo stars in velocity space exhibit abrupt changes at a
metallicity of [Fe/H] $\approx -1.7$~\citep{My18}. The metal-poor
population is isotropic, has prograde rotation $(\langle v_\phi
\rangle \approx 50$ km\,s$^{-1}$), mild radial anisotropy and a
roundish morphology (with axis ratio~$q\approx 0.9$).  
In contrast, the metal-rich
stellar population has almost no net rotation, is very radially
anisotropic and highly flattened with axis ratio~$q \approx 0.6 -
0.7$.


The velocity structure of the metal-rich population forms an elongated shape in velocity space, 
the so-called ``\emph{Gaia} Sausage''~\citep{Be18,MyGC}. 
It is believed to be caused by a substantial
recent merger~\citep{Be18,He18,Kr18}. The ``Sausage Galaxy'' must have collided
almost head-on with the nascent Milky Way to provide the abundance of
radially anisotropic stars. Even if its orbital plane was originally
inclined, dynamical friction dragged the satellite down into the
Galactic plane. Similarly, though its original orbit may only have
been moderately eccentric, the stripping process created tidal tails
that enforced radialisation of the orbit~\citep{Am17}, giving the
residue of highly eccentric stars in the \emph{Gaia}
Sausage. Therefore, the $\sim 10^{10}-10^{11}\Msol$ of DM in the
Sausage Galaxy~\citep{Be18,MyGC} will have been continuously stripped
over a swathe of Galactocentric radii, as the satellite sank and
disintegrated under the combined effects of dynamical friction and
radialisation.

The smooth round halo of the SHM cannot account for the highly
radially anisotropic DM associated with the \emph{Gaia} Sausage.
The SHM must therefore be extended to include a DM component with the radially
anisotropic kinematics that arise from the Sausage Galaxy merger. 
Before introducing our refinements in Section~\ref{sec:SHMpp},
we review the remaining ingredients of the SHM to discuss their validity.

\subsection{Sphericity}\label{sec:shape}
The stellar halo is clearly irregular as viewed in maps of resolved
halo stars~\cite{Belokurov:2006}. It comprises a hotchpotch of shells
and streams, many of which are associated with the \emph{Gaia} Sausage
(e.g.\ the Virgo Overdensity and the Hercules-Aquila Cloud~\cite{Si18}). 
However, analyses of the kinematic data from
\emph{Gaia} strongly suggest that the dark halo is a smoother and
rounder super-structure. Despite the abundant substructure, the
velocity ellipsoid of the stellar halo is closely aligned in spherical
polar coordinates~\citep{Ev16,We18}. This is a natural consequence of
the gravitational potential -- and hence the DM distribution -- being
close to spherical~\citep{Sm09,An16}.  Prior to data from \emph{Gaia},
there was a long-standing discrepancy regarding the dark halo
shape. Analyses of the kinematics of streams preferred almost
spherical or very weakly oblate shapes~\citep{Ko10,Bo15}. In contrast,
Jeans analyses of the kinematics of halo stars, which are subject to
substantial degeneracies between the stellar density, the velocity
anisotropy and the DM density, gave shapes varying from strongly
oblate to prolate~\citep{lo14,Bo16}. Reassuringly, the most recent
Jeans analyses of the kinematics of the stellar halo components with
\emph{Gaia} data release 2 (DR2) RR Lyraes find that the DM
distribution is nearly spherical~\citep{We18}, at least in the
innermost 15 kpc. This already suggests that the DM associated
with the \emph{Gaia} Sausage is subdominant. The DM halo must be a
smoother and rounder structure than the stellar halo. Therefore, the
assumption of near-sphericity in the potential that underlies the SHM
continues to be supported by the data.

\subsection{Circular Velocity at the Sun}\label{sec:v0}

The angular velocity of the Sun, derived from the Very-long-baseline
interferometry proper motion of Sgr A$^\star$ assuming it is at rest
at the centre of the Galaxy, is known accurately as $30.24 \pm 0.12$
km s$^{-1}$ kpc$^{-1}$~\cite{Re04,Bl16}. Thanks to results from
the GRAVITY collaboration ~\citep{Ab18}, the solar position is pinned
firmly down as $8.122\pm 0.031$ kpc.  This corresponds to a tangential
velocity of $246 \pm 1$ km s$^{-1}$. The circular velocity of the
Local Standard of Rest is extracted by correcting for the Solar
peculiar motion and for any streaming velocity induced by the Galactic
bar. The former is known accurately thanks to careful modelling as
$(U,V,W) = (11.1\pm1.5, 12.2\pm2, 7.3\pm1)$ km\,s$^{-1}$ from
Refs.~\cite{Schonrich:2010,McMillan:2017}, whilst the latter is harder
to estimate but is likely close to zero~\citep{Bl16}. This gives the
Local Standard of Rest as $v_0 = 233 \pm 3 \kms$.

Most direct detection experiments analyse their results 
with $v_0 = 220$ kms$^{-1}$, for recent examples, see analyses by the
SuperCDMS~\cite{Agnese:2014aze}, XENON~\citep{Aprile:2018dbl},
LUX~\citep{Ak17} and LZ~\citep{Akerib:2018lyp} Collaborations.
  Theoretical papers similarly continue to recommend $v_0 = 220
\pm 20$ kms$^{-1}$~\citep{Str13,Green:2017odb,Krauss18}. As a
consequence, the updated value $v_0 = 233~\kms$, together with its
substantially reduced error bar, are not currently a standard component in the
analysis of the experimental data.

\subsection{Escape Speed at the Sun}\label{sec:vesc}
The escape speed is directly related to the local potential, and
hence the mass of the Milky Way DM halo. Any revisions of the
escape speed are therefore important to include in refinements of the
SHM. 

Prior to \emph{Gaia}, measurements of the escape velocity relied on
radial velocities of small samples of high velocity stars.  For
example, the value of $\vesc = 544$ kms$^{-1}$ used in the
SuperCDMS~\cite{Agnese:2014aze}, XENON~\citep{Aprile:2018dbl},
LUX~\citep{Ak17} and LZ~\citep{Akerib:2018lyp} analyses is based on
the work of Ref.~\citep{Sm09}, who used a sample of 12 high velocity
RAVE stars. This was subsequently revised to $\vesc = 533^{+54}_{-41}$
kms$^{-1}$ when the sample size was increased to 90 stars~\cite{Pi14}. 
The escape speed curve as a function of Galactic
radius was measured in Ref.~\citep{Williams:2017} using a much larger
sample of $\sim 2000$ main-sequence turn-off, blue horizontal branch
and K giant stars extracted from the SDSS spectroscopic dataset.  The
local escape speed was found to be $521^{+46}_{-30}$ km s$^{-1}$.

However the proper motions in the \emph{Gaia} data enable a much
improved calculation, as we no longer need to marginalize over the
unknown tangential velocities of stars. Based on the analysis of the
velocities of $\sim 2850$ halo stars from \emph{Gaia} DR2 with
distance errors smaller than 10~\!\%, the local escape speed has been
revised upward to $\vesc = 580^{+63}_{-63}$ km s$^{-1}$~\citep{Mo18}.
However, Ref. \citep{Deason19} show that this result is sensitive to
the prior chosen to describe the high velocity tail of the
distribution function. Using a prior inspired by simulations, and a
more local sample of $\sim 2300$ high velocity counter-rotating stars,
they find the escape speed is $528_{-25}^{+24}$
kms$^{-1}$~\citep{Deason19}. This is compatible with the earlier work
of Refs.~\citep{Sm09,Williams:2017}, but with much smaller error bars.

\subsection{Local Dark Matter Density}\label{sec:rho0}

WIMP direct detection searches have traditionally taken $\rho_0=0.3$ GeV cm$^{-3}$
for the local DM density.
This is on the recommendation of the Particle Data Group Review~\citep{Am08},
although the works cited are not especially recent,
e.g.\ Ref.~\cite{Ga95}. On the other hand, axion haloscope
collaborations (ADMX~\cite{Duffy:2006aa,Asztalos:2009yp,Du:2018uak},
HAYSTAC~\cite{Brubaker:2018ebj,Brubaker:2017rna,Zhong:2017fot},
ORGAN~\cite{McAllister:2017lkb,McAllister:2017ern}) appear to have
independently decided on the value $\rho_0 = 0.45 \GeVcm$.

The consensus of recent investigations using the vertical kinematics
of stars tend to even larger values: in particular, $\rho_0 \approx
0.57$ GeV cm$^{-3}$ with Sloan Digital Sky Survey (SDSS) Stripe 82
dwarf stars~\cite{Smith:2012}; $0.542\pm0.042$ GeV cm$^{-3}$ with
4600 RAVE red clump
stars~\cite{Bienayme:2014}; $0.48 \pm 0.07$ GeV cm$^{-3}$ using a
model of the Galaxy built from 200,000 RAVE giants, together with
constraints from gas terminal velocities, maser observations and the
vertical stellar density profile~\citep{Pi14}; $0.46^{+0.07}_{-0.09}$
GeV cm$^{-3}$ with the SDSS G dwarfs~\cite{Sivertsson:2017rkp};
$0.69\pm0.08$ GeV cm$^{-3}$ with the Tycho \emph{Gaia} Astrometric
Solution (TGAS) red clump stars~\cite{Ha18}.  The statistical errors
on each of these measurements are smaller than the scatter between
them. This is because the error is dominated by systematics
(e.g. local gradient of the circular velocity curve, vertical density
law of disk tracers, treatment of the tilt of the velocity ellipsoid, see Ref.~\cite{Read:2014qva})
and probably amounts to $\approx 30\%$.

Fortunately $\rho_0$ only ever enters into calculations as an overall
scaling.  As a good basis for comparison between the work of different
groups, we suggest a suitable choice of rounded-off value for $\rho_0$
in the SHM$^{++}$ is 0.55 GeV cm$^{-3}$ with a $30 \%$ error of $\pm
0.17$ GeV cm$^{-3}$ to account for systematics.

\section{The SHM$^{++}$}
\label{sec:SHMpp}

\begin{figure*}[t]
\centering
  \includegraphics[width=0.49\textwidth]{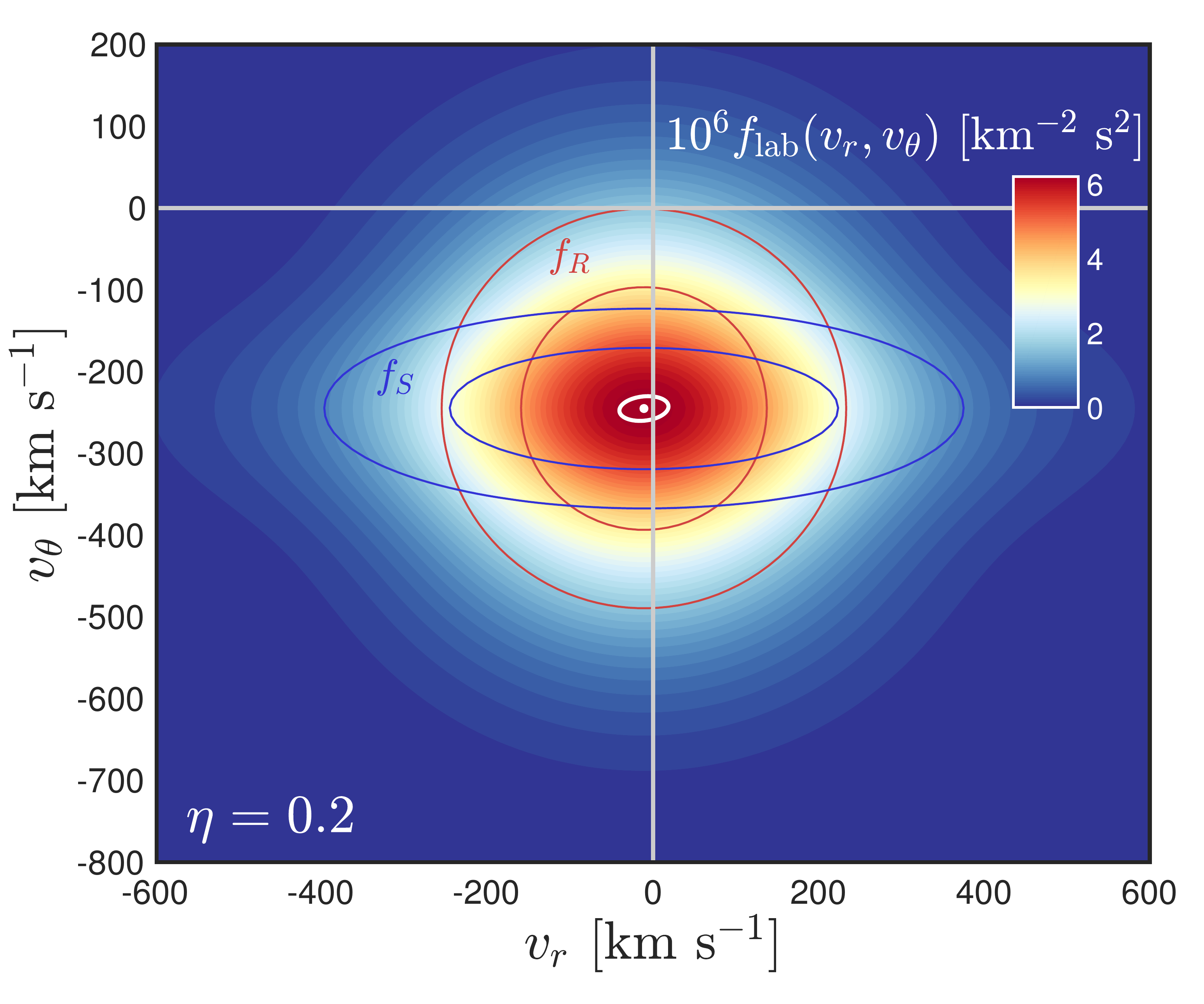}
  \includegraphics[width=0.49\textwidth]{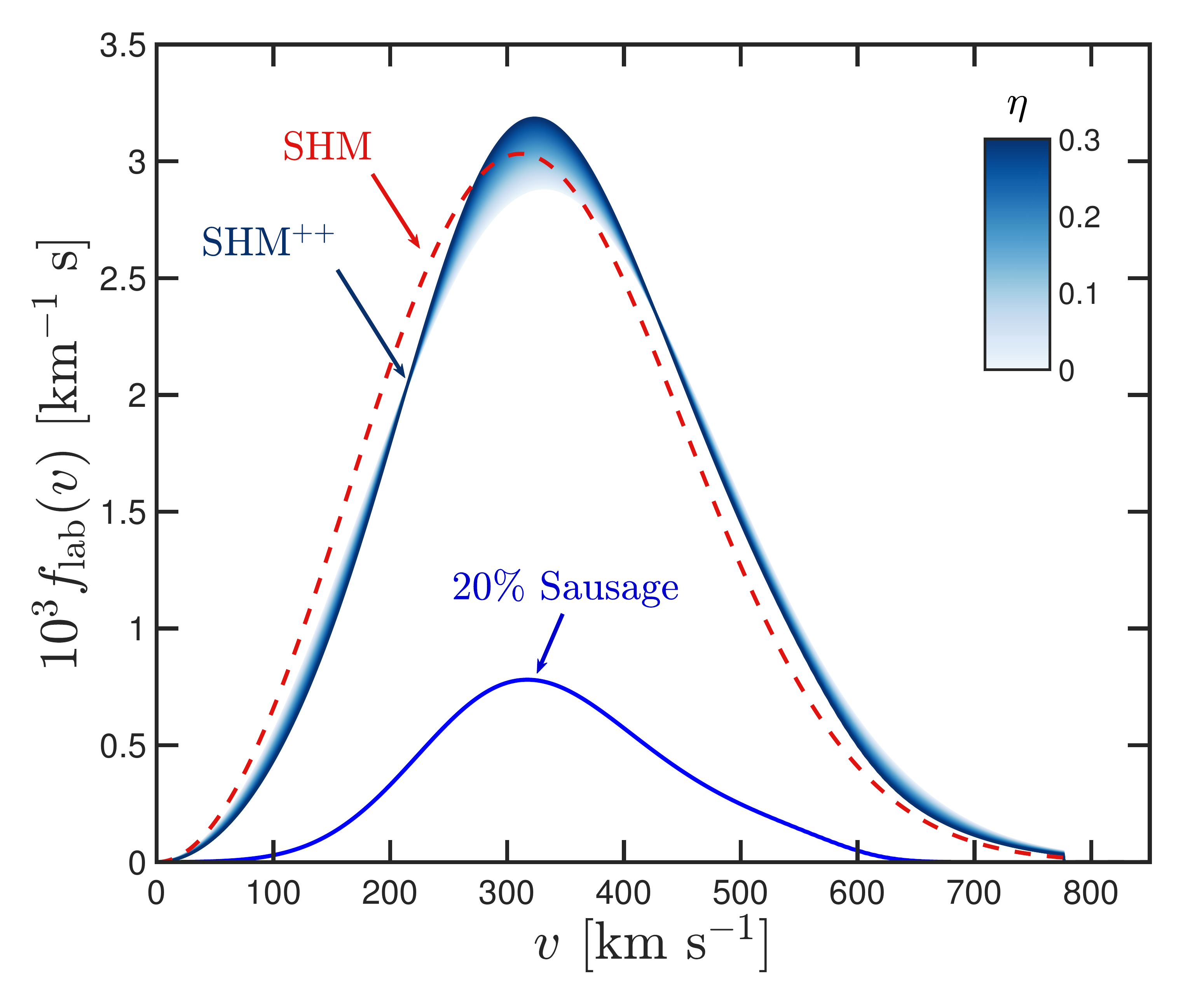}
\caption{{\bf Left:} Earth frame velocity distribution for the
  SHM$^{++}$ in the radial and horizontal directions. We assume a
  Sausage fraction of $\eta = 0.2$. 
  The shapes of the round component, $f_R(\mathbf{v})$, and Sausage component,
  $f_S(\mathbf{v})$, in velocity space are traced with red and blue
  contours respectively. 
  The radial anisotropy of the Sausage component can be clearly seen.
  The white point marks the inverse of the velocity of the Sun (LSR + peculiar motion)
   and the white circle indicates the path of
  the full Earth velocity over one year. {\bf Right:} Earth frame speed
  distributions for the SHM (red dashed) and the SHM$^{++}$ (blue). The
  shade of blue indicates the fraction of the halo comprised of
  Sausage. The lower blue line isolates only
  0.2$\fS(v)$. The effect of the Sausage component is to make the speed
  distribution colder. 
  }\label{fig:fv}
\end{figure*}

In this section, we introduce the SHM$^{++}$ by carrying out two
modifications to the SHM. First, and trivially, the local circular
speed $v_0$, escape velocity $\vesc$ and local DM density $\rho_0$ are
updated in the light of more recent data.  Second, and more
fundamentally, we introduce a Sausage component to describe the
radially anisotropic DM particles brought in by the Sausage galaxy.

\subsection{Velocity Distributions}

The velocity distribution of the bimodal dark halo in the frame of the
Galaxy is described by
\begin{equation}\label{eq:shmpp}
  f(\vbf) = (1- \eta) \fR (\vbf) + \eta \fS
  (\vbf) \, ,
\end{equation}
where $\fR$ is the velocity distribution of the smooth, nearly round
dark halo that dominates the gravitational potential in the innermost
20 kpc, whilst $\fS$ is the velocity distribution of the \emph{Gaia}
Sausage. The parameter $\eta$ is a constant that describes the
fraction of DM in the Sausage at the solar neighbourhood.

The nearly round dark halo component has a velocity distribution in
the Galactic frame that is the familiar Gaussian distribution in
Eq.~\eqref{eq:shm} with $v_0 = \sqrt{2} \sigma_v = 233\pm 3$
kms$^{-1}$ as the speed of the LSR. This relation holds true provided
the rotation curve is flat. The escape velocity used to cut off the
velocity distribution is $\vesc = 528^{+24}_{-25}$ km
s$^{-1}$~\citep{Deason19}.

We now turn our attention to a velocity distribution for the highly
radially anisotropic \emph{Gaia} Sausage. The velocity dispersion
tensor is aligned in spherical polar coordinates with $\sigma^2 = {\rm
  diag}(\sigma^2_r,\sigma^2_\theta, \sigma^2_\phi)$.\footnote{We use galactocentric spherical coordinates, which are equivalent to rectangular coordinates at the Earth's location.}  As the
gravitational potential is close to spherical~\cite{Ev16,We18}, then
$\sigma_\theta = \sigma_\phi$. The anisotropy is parameterized by,
\begin{equation}
  \beta = 1 - \frac{\sigma^2_\theta+\sigma_\phi^2}{2\sigma^2_r} \, ,
  \end{equation}
which vanishes for an isotropic dispersion tensor. We recall that
$\beta =1$ implies that all the orbits are completely radial and
$\beta = -\infty$ that all the orbits are circular. The stellar debris
associated with the \emph{Gaia} Sausage has $\beta =
0.9$~\cite{Be18,My18}. The Globular Clusters once associated with the
Sausage Galaxy are in fact even more radially anisotropic with $\beta
=0.95$~\cite{MyGC}. The anisotropy of the Sausage DM is unknown,
though it too must be highly radial. We assume it is the same as the
stellar debris $\beta = 0.9$ in our standard model and assign an error of~$\pm 0.05$.

The density distribution of the Sausage falls like~$\sim r^{-3}$~\citep{Io18}.
 The exact solution of the collisionless Boltzmann
equation for an anisotropic tracer population with density falling
like $r^{-3}$ in a galaxy with an asymptotically flat rotation
curve is~\citep{Ev97}.
\begin{align}\label{eq:sausage}
\fS(\vbf) =
\frac{1}{(2\pi)^{3/2}\sigma_r\sigma_\theta^2 N_\mathrm{S,esc}}
\, \exp \left( - \frac{v_r^2}{2\sigma_r^2}
-\frac{v_\theta^2}{2\sigma_\theta^2}- \frac{v_\phi^2}{2\sigma_\phi^2}
\right) \nonumber\\ \times \Theta (\vesc - |\vbf|)\,,
\end{align}
The velocity dispersions are related to the amplitude of the rotation
curve via~\citep{Ev97}
\begin{equation}\label{eq:disps}
  \sigma_r^2 = \frac{3v_0^2}{2(3-2\beta)},\qquad\sigma_\theta^2 =
  \sigma_\phi^2 = \frac{3 v_0^2(1-\beta)}{2(3-2\beta)}\,,
\end{equation}
where $v_0 =233$~kms$^{-1}$ is the LSR.

The normalisation constant is
\begin{align}
N_\mathrm{S,esc} = \erf \left( \frac{\vesc}{\sqrt{2}\sigma_r}\right) -
\left(\frac{1-\beta}{\beta}\right)^{1/2}
\exp \left(-\frac{\vesc^2}{2\sigma_\theta^2} \right)\nonumber\\
\times \erfi \left( \frac{\vesc}{\sqrt{2}\sigma_r}\frac{\beta^{1/2}}{(1-\beta)^{1/2}}\right),
\end{align}
where $\erfi$ is the imaginary error function. This is the anisotropic
analogue of Eq.~(\ref{eq:norm}).

This completes our description of the velocity distribution of the
SHM$^{++}$. It is an {\it entirely analytic model} of a roundish dark
halo, together with a highly radially anisotropic Sausage
component. It depends on the familiar Galactic constants already
present in the SHM, namely the local circular speed $v_0$, the local
escape speed $\vesc$ and the local DM density $\rho_0$. There are two
additional parameters in the SHM$^{++}$:  the velocity
anisotropy $\beta \approx 0.9 \pm 0.05$ of the \emph{Gaia} Sausage
and the fraction of DM locally in the Sausage $\eta$, which
we estimate in the next section.

On Earth, the incoming distribution of DM particles is found by boosting
the DM velocities in the galactic frame by the Earth's 
velocity with respect
to the Galactic frame: ~$\mathbf{v}_{\mathrm{E}}(t)=(0,v_0,0)+(U,V,W)+\mathbf{u}_{\mathrm{E}}(t)$.
Explicitly, this means that the Earth frame velocity distribution is
$f_{\rm{lab}}(\mathbf{v})=f(\mathbf{v}+\mathbf{v}_{\mathrm{E}}(t))$.
The Earth's velocity is time dependent
owing to the time dependence of $\mathbf{u}_{\mathrm{E}}(t)$,
the Earth's velocity around the Sun.  Expressions for
$\mathbf{u}_{\mathrm{E}}(t)$ are given in~Refs.~\cite{Lee:2013xxa,
  McCabe:2013kea,Mayet:2016zxu}.
  
We plot the Earth frame distribution of
velocities and speeds in Fig.~\ref{fig:fv}.
The velocity distribution (left panel) is
displayed as the two-dimensional distribution $f_{\rm{lab}}(v_r,v_\theta)$,
 where we have marginalised over~$v_\phi$. 
 The blue contours associated with the Sausage component clearly show the
 radial bias in velocity space compared to the circular
 red contours associated with the round component of the halo.
In the right panel, we show the speed distribution,
$f_{\rm{lab}}(v)=v^2 \int d \Omega f_{\rm{lab}}(\mathbf{v})$,
for the SHM, SHM$^{++}$ and the isolated Sausage component.
For the SHM distribution (red dashed line), we have used the
parameters in the upper half of Table~\ref{tab:astrobenchmarks}.
For the SHM$^{++}$ distribution (blue shaded), we have used the parameters in
the lower half of Table~\ref{tab:astrobenchmarks} with the exception of~$\eta$, which
we have allowed to vary in the range $\eta=0$ 
(corresponding to only a round halo component)
to $\eta=0.3$. The solid blue line shows the contribution from only the Sausage
component with $\eta=0.2$.

Comparing the SHM and SHM$^{++}$ distributions, we see that the
SHM$^{++}$ distribution is everywhere shifted to higher speeds. This
is primarily because of the larger value of $v_0$.  Comparing the
SHM$^{++}$ distribution with $\eta=0$ (the lightest edge in the shaded
region) to the distribution with $\eta\neq0$, we see that the impact
of the Sausage component is to increase the peak-height of the speed
distribution while decreasing the overall dispersion of the
distribution, i.e.\ the Sausage component makes the total speed
distribution colder compared to a halo with only the round, isotropic
component.  The difference in the dispersion arises from the different
expressions for the velocity dispersions in the Sausage distribution
($\fS$) compared to the round halo~($\fR$).

\subsection{Constraining $\eta$}
The fraction $\eta$ of DM locally in the \emph{Gaia} Sausage is not
known, but an upper limit can be estimated. The stellar
density distribution of the Sausage is triaxial with axis ratios $a =1
$, $b = 1.27\pm 0.03$, $c = 0.57 \pm 0.02$ near the Sun, and falls off
like $\sim r^{-3}$~\citep{Io18}. As a simple model, we assume that the
Sausage DM density is stratified on similar concentric ellipsoids with
ellipsoidal radius $m$
\begin{equation}
    m^2 = {x'^2\over a^2} + {y'^2\over b^2} + {z^2\over c^2}.
\end{equation}
Here, ($x',y'$) are the Cartesians in the Galactic plane, rotated so
that the long axis $x'$ is about $70^\circ$ with respect to the
$x$-axis which conventionally connects the Sun and the Galactic
Centre~\citep{Io18}.

The DM contribution of the triaxial Sausage cannot become too high, as
it would then cause detectable perturbations (in the rotation curve or
the kinematics of stars, for example) and would spoil the sphericity
of the potential~\citep{Ev16,We18}.  For large spirals like the Milky
Way, the scatter in the Tully-Fisher relationship severely limits the
ellipticity of the disk~\citep{Fr92}.  In fact, the ellipticity of the
equipotentials in the Galactic plane of the Milky Way must be less
than 5 \% on stellar kinematical grounds~\citep{Ku94}, almost
all of which can be attributed to the Galactic bar~\cite{Mu03}. Any
contribution to the ellipticity of the equipotentials in the Galactic
plane from the Sausage must be less than $\sim 1 \%$.

To estimate the dynamical effects of the Sausage, we need to compute
the gravitational forces generated by an elongated, triaxial figure.
For now we assume that the DM density falls in the same manner as the stars
 so the Sausage density within $m <30$ kpc is modelled by
\begin{equation}\label{eq:rhosos}
  \rhoS(m^2) = {\rho_a r_a^3\over (r_a^2+m^2)^{3/2}}.
\end{equation}
The virtue of this model is that the gravitational potential of the
Sausage at any point is then known~\citep{Chandra87}
\begin{equation}
  \phi = -4\pi G \rho_a abc R_{\rm F}(\lambda,\mu,\nu)\;,
\end{equation}
where $R_{\rm F}$ is the Carlson elliptic integral and
($\lambda,\mu,\nu$) are ellipsoidal coordinates.  The total mass
within ellipsoidal radius $m$ is
\begin{align}\label{eq:Mtot}
  M(m) = 4\pi a b c \, r_a^3\,\rho_a\Bigl[ \log( [m +\sqrt{r_a^2+m^2}] /r_a)\\
       -{m\over
      \sqrt{r_a^2+m^2}} \Bigr]\nonumber\;.
\end{align}
Although the total mass diverges logarithmically, this is not a
problem as by Newton's theorem, ellipsoidal shells of constant density
have no dynamical effects inside the shell.

We now constrain the mass of the Sausage within 30~kpc, or
$M(30)$. The Sausage contributes a monopole component which provides a
small part of the local circular velocity speed of $v_0 = 233$
kms$^{-1}$.  The remainder is provided by the rest of the Galaxy. This
is modelled as a logarithmic potential with an amplitude chosen so
that, when its circular speed is added in quadrature to that of the
Sausage, the local circular speed of 233 kms$^{-1}$ is correctly
reproduced.  Requiring the ellipticity of the combined equipotentials
in the plane to be less than 1\% imposes an upper limit on the
mass of the Sausage~$M(30)$ of $\lesssim 3 \times 10^{10}
M_\odot$. Using the density law Eq.~\eqref{eq:rhosos}, we find the
fraction of DM in the Solar neighbourhood due to the Sausage is $\eta
\lesssim 20\%$.

There are arguments suggesting that this limit may be an overestimate
-- for example, the DM is always more extended than the luminous
matter in dwarf galaxies. Tidal stripping of an infalling satellite
therefore distributes DM over a much large volume than the luminous
matter, so our use of $\sim 30$ kpc inspired by the stellar debris may
be unwarranted. The density law of the DM may also be different from
the $r^{-3}$ fall-off of the stars.  However, there are also arguments
suggesting that this may be an underestimate -- for example, the
velocity distribution of the stellar debris~\cite{La18} suggests that
the stellar density is depleted in the very innermost parts. If the
same is true of the DM, then our calculation may not ascribe enough DM
to the solar neighbourhood. Given all the uncertainties, it is
therefore prudent to allow $\eta$ to vary within the range $10\%$ to
$30\%$ with a preferred value of $20\%$. This is consistent with
recent numerical work from the Auriga~\citep{Fa18} and the FIRE
simulations~\citep{Fire18}.

\section{Comparisons with Alternatives}\label{sec:alternatives}

Our derivation of the SHM$^{++}$ is based upon equilibrium
distribution functions, together with a collection of robust
astronomical measurements. The main advantage of our model is that it
is both simple and accounts for known properties of the Milky Way halo. 

However, it is not the only method that has been used to infer the
local DM velocity distribution.  Recently, there have been attempts to
deduce the velocity distribution empirically using observations of
low metallicity halo stars. We can also resort to numerical simulations to gain
more understanding of the behaviour of DM inside galactic halos.  We
discuss possible overlaps and disagreements with these various methods
here.

\subsection{Low and Intermediate metallicity halo stars}

An alternative suggestion as to an appropriate velocity distribution
is motivated by the claim that the metal-poor halo stars are effective
tracers of the local DM
distribution~\citep{Herzog-Arbeitman:2017fte}. This claim has inspired
DM velocity distributions based on the empirical properties of the
velocities of the metal-poor stars in RAVE or
SDSS-\emph{Gaia}~\citep{Herzog-Arbeitman:2017zbm, Necib:2018iwb}.
However, if the velocity distributions of any two populations are the
same, and they reside in the same gravitational potential, then their
orbital properties are also the same. In gravitational physics, the
density of stars or DM is built up from their orbits. So, the
assumption is equivalent to assuming that the density of the stars and
DM are the same.  This remains true even if the potential is not
steady.

Hence, the hypothesis of
Refs.~\cite{Herzog-Arbeitman:2017fte,Herzog-Arbeitman:2017zbm,
  Necib:2018iwb} is equivalent to assuming that the density
distributions of the metal-poor halo stars and the DM are the same (up
to an overall normalization). This is known to be incorrect, as the
density of the metal-poor stars (or any stellar component) falls too
quickly with Galactocentric radius to provide the flatness of the
Milky Way's rotation curve.
This causes the local velocity distribution of low metallicity halo
stars to be much colder than the DM. Consequently, if the DM velocity
distribution is assumed to follow the metal-poor stars, then the dark
matter speeds will be under-estimated (as their orbits no longer
provide the density at large radii to make the rotation curve
flat). This leads to the conclusion that the DM is colder than is
really the case.  Our argument is corroborated by results from the
Auriga simulations~\cite{Grand:2016mgo}, where the speed profiles of
low metallicity stars in the simulated halos are indeed colder than
the DM speed distributions~\cite{BozorgniaAuriga:2018}.

Of course, the DM may have multiple sub-populations, some of which
track the stellar density distributions and some of which do not. In
fact, this is seen in some of the insightful examples provided in the
FIRE simulations~\cite{Fire18}. Then of course the density of stars
and DM can be different. However, if only a minority of the DM tracks
the stellar density, then the analogy is of only partial help in
providing velocity distributions for direct detection experiments.

In the picture of Ref.~\cite{Necib:2018iwb}, the correspondence
between the metal-poor stars and dark matter pertains only to the
oldest luminous mergers that build the smooth, nearly round dark halo
(our $\fR$).  The Sausage stars are of intermediate metallicity, and
here the assumption is that they trace the DM brought in by the merger
event.  The humped structure in $f(v_r)$ with two lobes at $v_r = \pm
148$ km s$^{-1}$ is associated with the apocentric pile-up at $\sim
30$ kpc that marks the density break in the stellar
halo~\cite{Lancaster:2018}.  They will not exist for the Sausage DM
velocity distribution, as the DM density profile does not break at
$\sim 30$~kpc.  The DM from the Sausage Galaxy was originally more
extended than the stars in the progenitor and so was stripped earlier
and is likely sprawled over much larger distances.  In simulations of
sinking and radializing satellites, the length scale of the DM tidally
torn from the satellite exceeds that of the stars by typically a
factor of a few ~\cite[e.g.,][]{Am17,Fa18}. The density of the tidally
stripped stars and DM from the Sausage Galaxy will also therefore be
quite different.

\subsection{Simulations}
Our data-driven work is complementary to the approach
adopted in Refs.~\cite{Bozorgnia:2016ogo,Kelso:2016qqj,Sloane:2016kyi,Lentz:2017aay}. 
Here, simulated halos built up from successive merger events are examined to
extract better motivated velocity distributions than the SHM
ansatz. In part, this is also an attempt to understand the connections
present between the dark and baryonic matter distributions
of galactic halos. A detailed summary of the findings of a collection of 
simulations and their implications for direct detection can be found
in Ref.~\cite{Bozorgnia:2017brl}.

Several of these studies confirm that the Maxwellian speed
distribution derived from the SHM is satisfactory for the purposes of
direct detection signal modelling.
Refs.~\cite{Bozorgnia:2016ogo,Bozorgnia:2017brl} raise the caveat that
in some simulations, the circular speed $v_0$ and peak speed of the
distributions are different, though this was not found
in~\cite{Kelso:2016qqj}.

A key difference in our approach is that we have made a bespoke
velocity distribution to account for a known merger event in the Milky
Way's recent history. In the future, the complementarity between our
approach and numerical simulations will grow, as we can use
simulations to understand more about the impact of this merger on the
local DM distribution. Ultimately this will put both data-driven and
simulation-driven predictions of experimental DM signals on more
robust grounds.

\section{Experimental consequences}\label{sec:dm}

\begin{figure*}[t]
\includegraphics[width=0.49\textwidth]{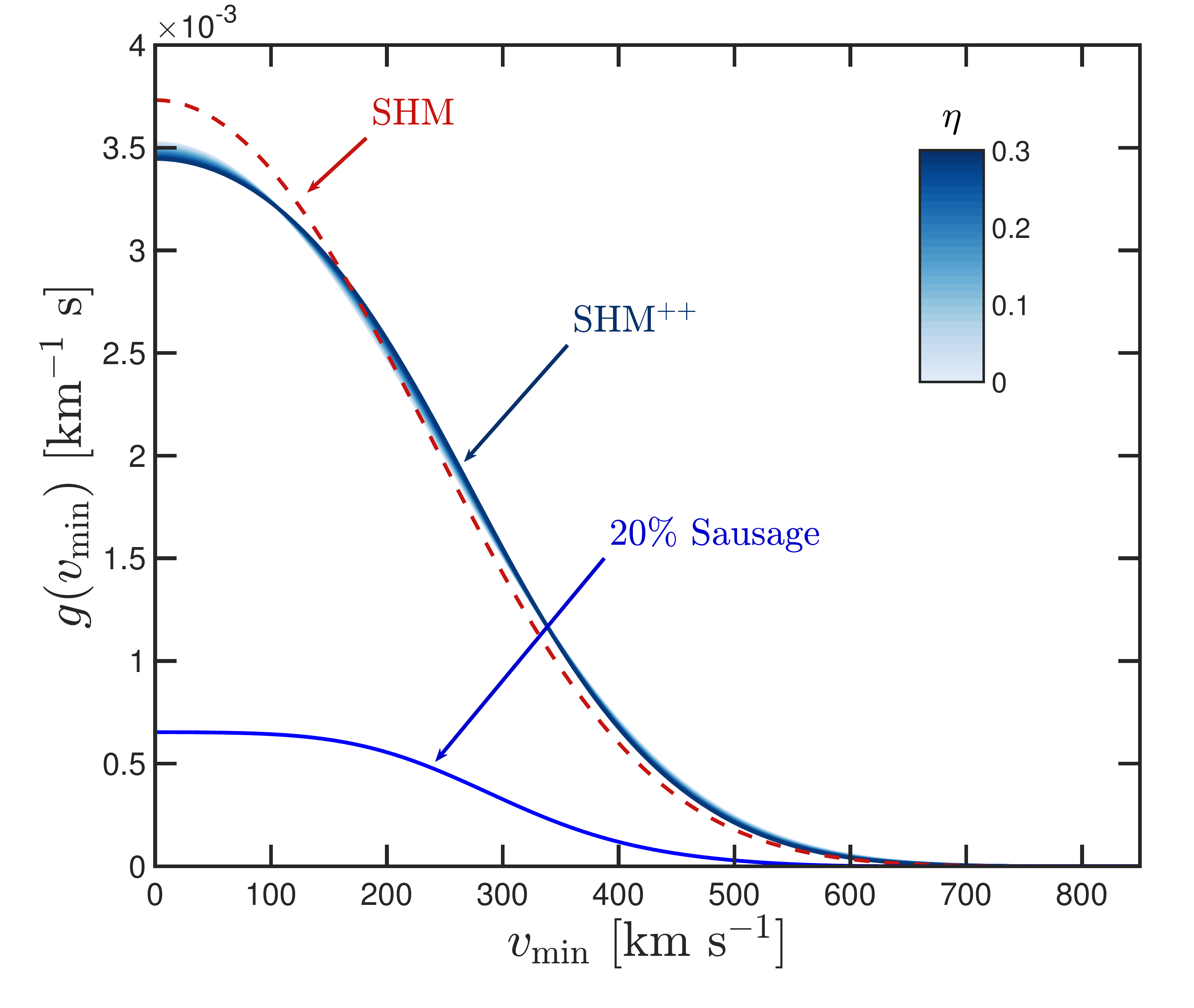}
\includegraphics[width=0.49\textwidth]{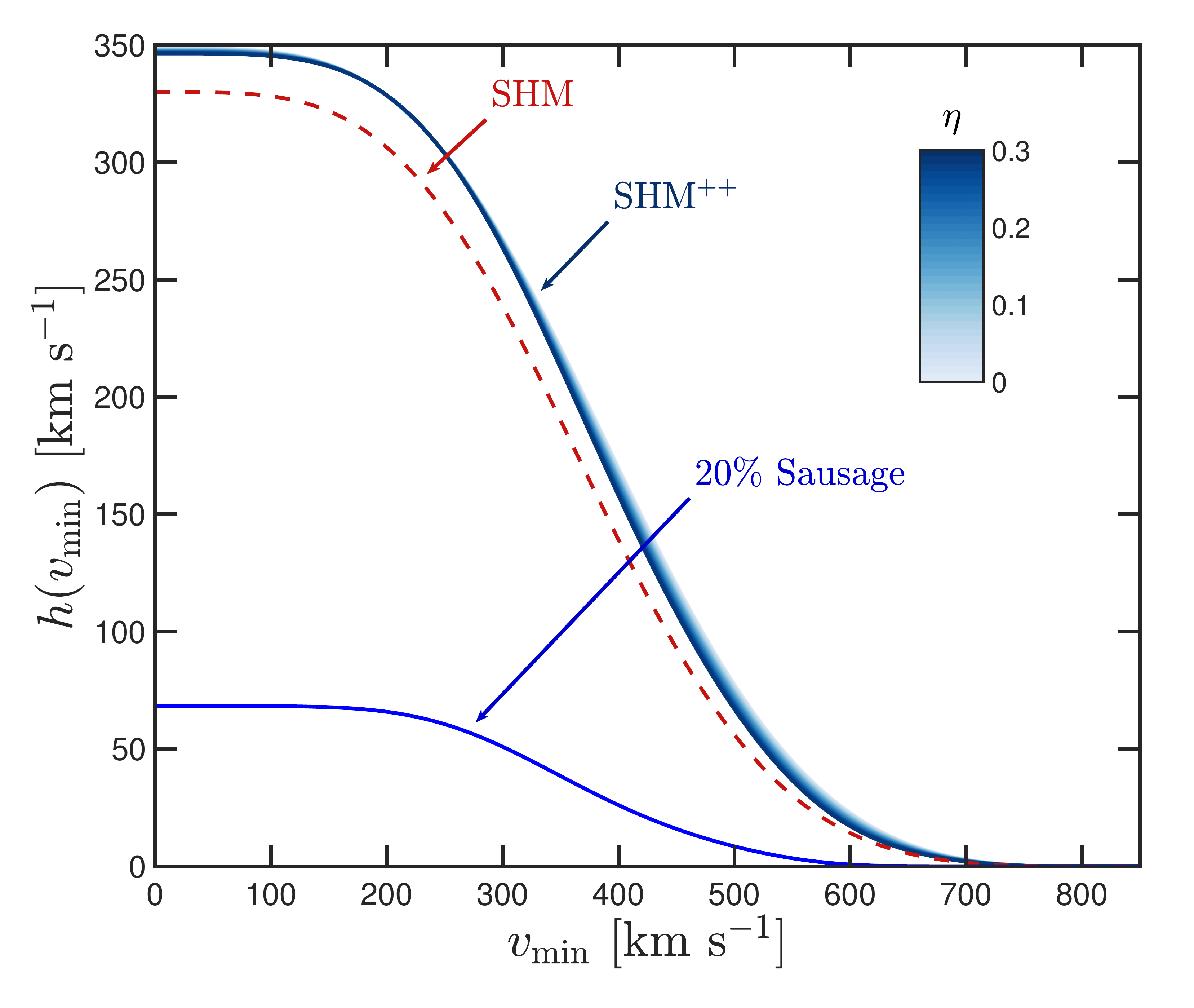}
\caption{The two halo integrals outlined in the text that enter the rate calculation
for nuclear recoil signals: the mean inverse
  speed $g(\vmin)$ ({\bf left}) and mean speed $h(\vmin)$ ({\bf right}). We show
  their shapes for both the SHM and SHM$^{++}$ as a red line and blue
  region respectively. We take the value of these functions averaged
  over time.  Direct detection event rates are linearly
  proportional to one or both of these integrals. The blue shading
  indicates the value of the Sausage density fraction,
  $\eta$. The Sausage component with~$\eta=0.2$ is isolated as a blue
  line.}\label{fig:gvmin}
\end{figure*}

The purpose of a standard benchmark halo model is to facilitate the
self-consistent mapping of exclusion limits on the properties of a
DM particle candidate. Since all detection signals require
this input, all are influenced by changing the model of the
local distribution of DM. In this section, we demonstrate in simple terms the
differences brought about by the bimodal distribution of the \SHMpp. As
examples, we consider the two most popular candidates for DM:
weakly interacting massive particles (WIMPs) and axions. 
For WIMPs, we show the effect of the Sausage on 
nuclear recoil event rates
(Sec.~\ref{sec:recoils}) and cross section limits
(Sec.~\ref{sec:limits}), as well as differences in their
directional signals (Sec.~\ref{sec:directional}). For
axions (Sec.~\ref{sec:axions}), the discussion is more straightforward
and can be summarised with a simple formula encapsulating the
consequences of different signal model assumptions.

\subsection{Nuclear recoil signals}\label{sec:recoils}

There are many experiments actively searching for the nuclear recoil
energy imparted by the collision of a DM particle with a
nucleus.  For these two-to-two scattering processes (which could be
elastic or inelastic collisions~\cite{TuckerSmith:2001hy,Baudis:2013bba,McCabe:2015eia}), 
the general formula for the
differential scattering rate~$R$ of nuclear recoil events as a
function of the nuclear recoil energy~$E_r$ is
\begin{equation}\label{eq:eventrate}
 \frac{\textrm{d}R(t)}{\textrm{d}E_r} = N_T \frac{\rho_0}{m_\chi} \int_{v>\vmin} \!\!\!\!  v \, f\big(\mathbf{v}+\mathbf{v}_{\mathrm{E}}(t)\big) \,\frac{d \sigma_T (v,E_r)}{d E_r}  \, \textrm{d}^3v \, .
\end{equation}
Here, $N_T$ is the number of target nuclei in the experiment,
$m_{\chi}$ is the DM mass, $v=|\mathbf{v}|$ is the DM speed in the reference frame of the experiment, $v_{\rm{min}}$ is the minimum DM speed
that can induce a recoil of energy~$E_r$, $\sigma_T$ is the
DM--nucleus scattering cross section, which in general depends on $v$
and $E_r$, and finally $ f(\mathbf{v+v_{\mathrm{E}}}(t))$ is the DM velocity
distribution boosted to the Earth's frame.

For the canonical leading order spin-independent (SI) and
spin-dependent (SD) DM--nucleus interactions~\cite{Undagoitia:2015gya}, the differential
cross section is inversely proportional to the square of the DM speed,
$d\sigma_T/dE_r\propto v^{-2}$. For these interactions, all of the
dependence on the DM velocity distribution is encapsulated in the
function
 \begin{equation}\label{eq:gvmina}
g(v_{\rm min},t) = \int_{v>\vmin} \frac{f(\textbf{v}+\mathbf{v}_{\mathrm{E}}(t))}{v} \,
\textrm{d}^3 v \, .
 \end{equation}
More general two-to-two DM--nucleus interactions can be parameterized
in the non-relativistic effective field theory framework for direct
detection~\cite{Fitzpatrick:2012ix,Fitzpatrick:2012ib,Anand:2013yka},
which allows for any Galilean invariant and Hermitian interaction that
respects energy and momentum conservation. Within the effective field
theory framework, direct detection signals depend on a linear
combination of $g(v_{\rm min},t) $ and $h(v_{\rm
  min},t)$,\footnote{Scattering processes that are not two-to-two can
  have a more general velocity dependence that isn't captured by the
  non-relativistic effective field theory framework, see
  e.g.~\cite{Kouvaris:2016afs,McCabe:2017rln}.} which is defined as
 \begin{equation}\label{eq:gvminb}
h(v_{\rm min},t) = \int_{v>\vmin} v f(\textbf{v}+\mathbf{v}_{\mathrm{E}}(t)) \,
\textrm{d}^3 v \, .
 \end{equation}

We show $g(v_{\rm min},t) $ and $h(v_{\rm min},t)$ in
Fig.~\ref{fig:gvmin}. For the SHM, there exist analytic expressions
for these integrals (see
e.g.~\cite{Savage:2006qr,McCabe:2010zh,Fitzpatrick:2010br}).  For the
\SHMpp, there are no known analytic expressions, though they are
easily evaluated numerically.\footnote{We have provided a public code
to generate these functions. The code is available at \url{https://github.com/mccabech/SHMpp/}.}
 The blue shaded region
corresponds to the \SHMpp~with the Sausage fraction in the range $0.1
\le \eta \le 0.3$. The dashed red line shows the result for the SHM
with the parameters in Tab~\ref{tab:astrobenchmarks}.  For the mean
inverse speed, $g(\vmin)$, the \SHMpp~produces a slightly different
shape leading to a suppression of around 10\% for $\vmin<200 \kms$ and
a much smaller increase over speeds $\vmin>200 \kms$. The inclusion of
the Sausage component leads to a small change in the shape of the mean
speed integral, $h(\vmin)$, but there is a persistent increase
of around 6\% in the \SHMpp~relative to the SHM. This increase
reflects the 6\% increase in $v_0$ in the \SHMpp.

At first sight, it may seen surprising that the differences between
the SHM and \SHMpp~models are not greater. The resolution lies in the
fact that we have made multiple counter-balancing changes to the
SHM. The relative coldness of the new Sausage DM is counteracted by the
increased hotness of the halo DM due to the increase in $v_0$.

With $g(v_{\rm min},t)$ in hand, we compare the rate and exclusion 
limits in the SHM and SHM$^{++}$ for the most commonly studied interaction: 
the spin-independent DM-nucleus interaction, in which the differential cross section takes the
form:
\begin{equation}
\frac{d \sigma_T (v,E_r)}{d E_r} = \frac{m_N A^2 \sigma_p^{\rm{SI}}}{2 \mu_p^2\,v^2} F^2(E_r)\;,
\end{equation}
where $m_N$ is the nucleus mass, $A$ is the atomic number, $\mu_p$ 
is the DM-proton reduced mass, $F(E_r)$ is the nuclear form factor
and $\sigma_p^{\rm{SI}}$ is the DM-proton scattering cross section.

\begin{figure}[t]
\includegraphics[width=0.49\textwidth]{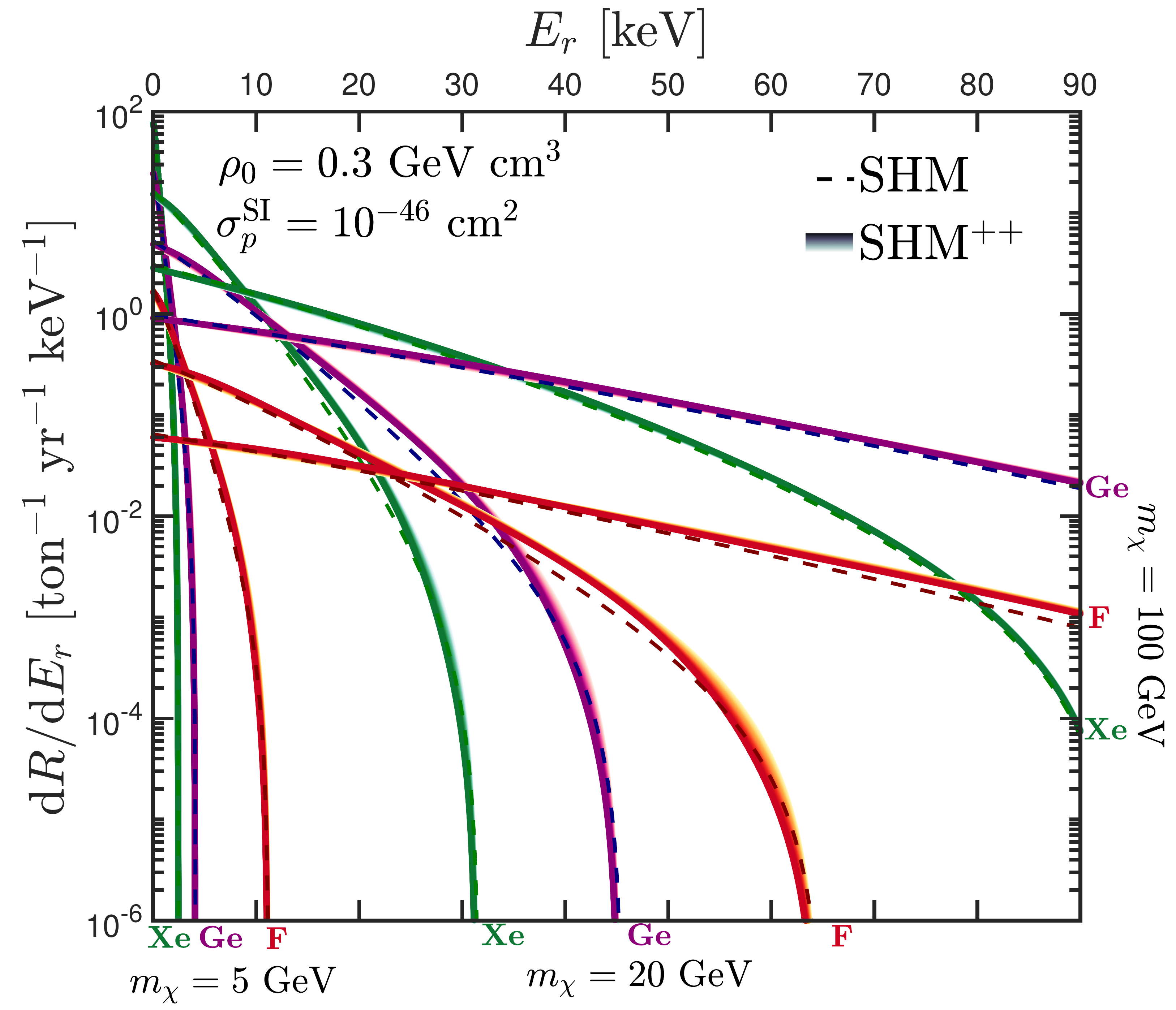}
\caption{Spin-independent differential event rate as a function of
  energy for the SHM (dashed) and SHM$^{++}$ (shaded, indicating a
  range of $\eta$, as in Figs.~\ref{fig:fv} and~\ref{fig:gvmin}).  The
  rates for three target nuclei are shown: xenon (green), germanium
  (purple) and fluorine (red).  We also show results for three
  different values of the DM mass (5, 20 and 100 GeV for rates
  extending from the lowest to the highest energies shown).  We have
  fixed $\rho_0=0.3$~GeV~\!cm$^3$ for the SHM and SHM$^{++}$ spectra
  to show only the changes due to the velocity distributions.
}\label{fig:eventrates}
\end{figure}

\begin{figure}[t]
\includegraphics[width=0.49\textwidth]{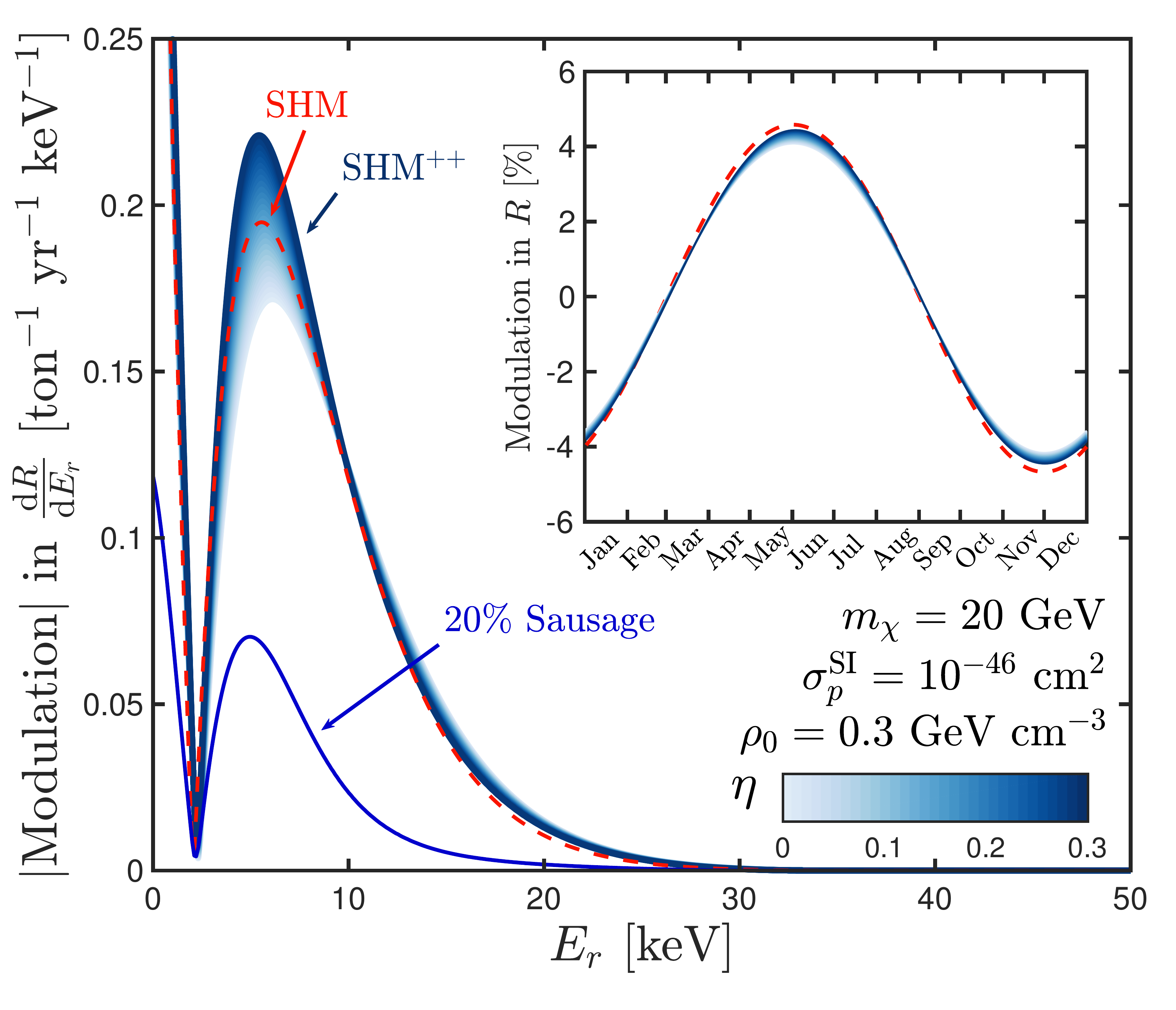}
\caption{Annual modulation of the SI differential event rate ({\bf
    main}) and total rate ({\bf inset}) for DM with a mass of 20~GeV
  scattering off a xenon nucleus. The blue shaded region corresponds
  to the SHM$^{++}$ with varying $\eta$ whereas the dashed red line is
  the SHM. As in Fig.~\ref{fig:eventrates}, we have fixed $\rho_0 =
  0.3 \GeVcm$ for both the SHM and SHM$^{++}$ to again isolate the
  changes brought about by the velocity distributions.
}\label{fig:annualmodulation}
\end{figure}

We show the differential event rate as a function of recoil energy in
Fig.~\ref{fig:eventrates} for a three different target nuclei:
$^{131}$Xe (green), $^{74}$Ge (purple) and $^{19}$F (red). In this
figure, we have temporarily fixed the local DM density at $\rho_0 =
0.3\GeVcm$ for both models.  This is so that we can highlight only the
difference arising from the change in velocity distribution, rather
than the simple rescaling by the new value of $\rho_0$.  Three
different values of the DM mass are shown and as in previous figures,
the shading indicates the effect of the Sausage component.  Consistent
with the rather minor alterations seen in $g(v_{\rm min})$, there are
only modest changes in the recoil energy spectra, mainly in the
high-energy tails of the spectra.  The spectra with only a round halo,
corresponding to $\eta=0$, and with the updated astrophysical
parameters in the lower half of Table~\ref{tab:astrobenchmarks} are
shown by the lightest colour in the shaded region. We see that
increasing~$\eta$ slightly reduces the maximum energy for all cases.

The differential modulation event rate, defined as 
$dR/dE_r\large|_{\rm{max}}-dR/dE_r\large|_{\rm{min}}$,
is shown in the main panel of Fig.~\ref{fig:annualmodulation}. 
We assume here a DM mass of 20 GeV scattering with xenon.  
As in previous figures, the red-dashed line shows the rate for the SHM,
 the blue-shaded regions shows the rate for the SHM$^{++}$ for different
values of~$\eta$ and the blue line shows the contribution from the Sausage
component with $\eta=0.2$.
We have again fixed $\rho$ to be the same for the SHM and SHM$^{++}$ spectra.
As in the previous figures, the changes between the two models are relatively small.
Increasing the contribution of the Sausage component has the effect of increasing 
the peak modulation amplitude while slightly decreasing the higher energy modulation spectrum.

The inset in Fig.~\ref{fig:annualmodulation} shows the modulation in the total scattering rate, $R$,
over the course of one year. Importantly, we note that the Sausage
component of the halo is modulating essentially in phase with the
isotropic part. Recall that the modulation of the event rate is
controlled by relative velocity between the motion of the Earth and
the direction in velocity space that the distribution is
boosted. Since both the halo and the Sausage are centred at the
origin in velocity space, they are both boosted to the same new centre in the Earth frame (see Fig.~\ref{fig:fv}). In fact it
is only features that are off-centred in velocity space,
such as streams, that can give
rise to significant phase changes in the annual modulation signal~\cite{Savage:2006qr,OHare:2018trr}. 

\subsection{Impact on cross section limits}\label{sec:limits}
 \begin{figure}[t]
\centering
\includegraphics[width=0.49\textwidth]{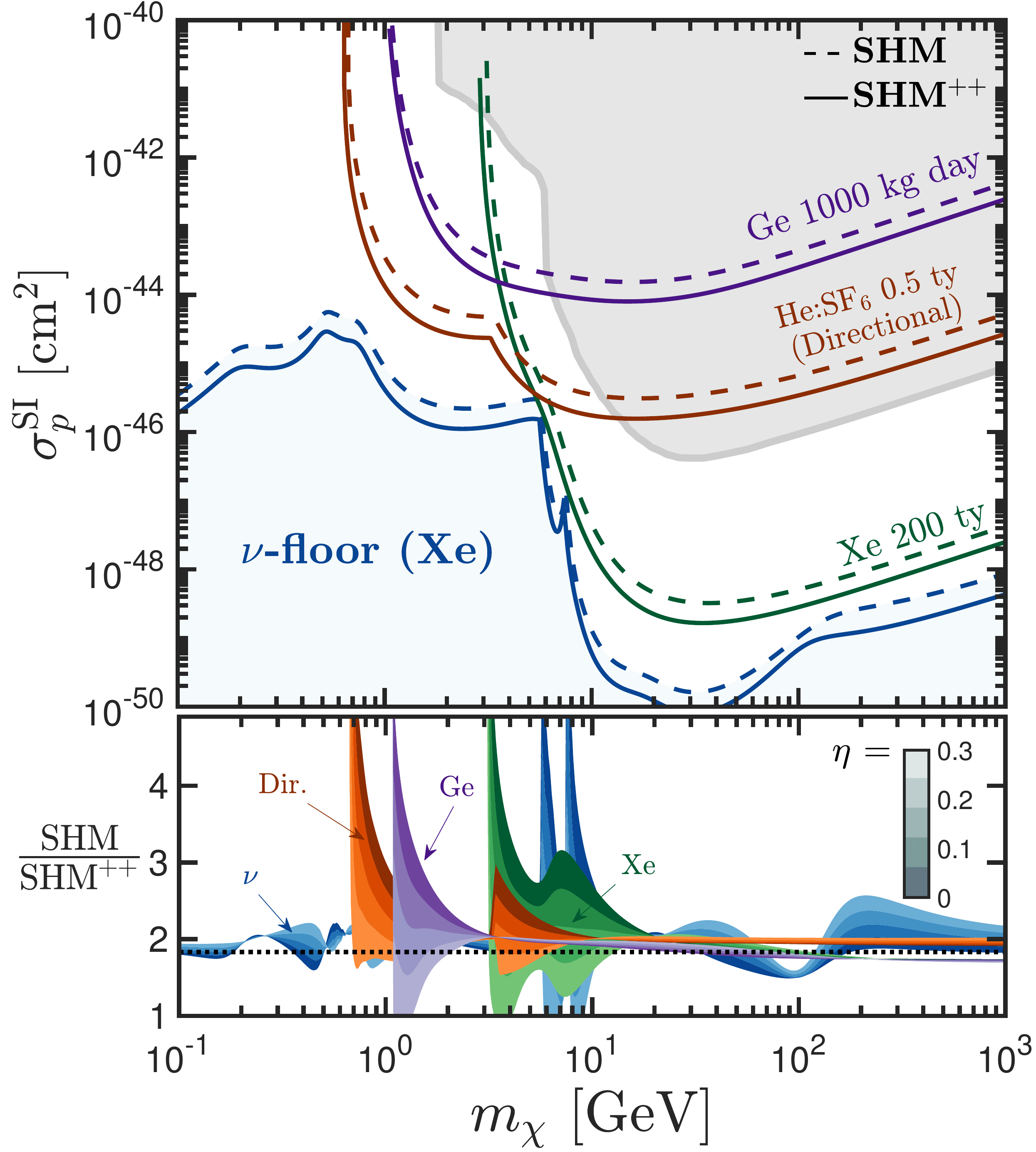}
\caption{{\bf Top:} Using a set of toy experimental setups, we
  demonstrate the impact of the SHM$^{++}$ on the sensitivity
  limits for three classes of detectors: a germanium experiment (purple), 
  a directional He:SF$_6$ experiment (orange) and a xenon experiment (green).
  The lower blue shaded region shows the neutrino
  floor for a xenon target while the grey shaded region shows the 
  already excluded parameter space (assuming the SHM).
The dashed lines indicate the sensitivity assuming
  the SHM while the solid lines assume the SHM$^{++}$.
 For the SHM$^{++}$ limits in the top panel, we have used 
 the parameters from the lower half of Table~\ref{tab:astrobenchmarks}.
  {\bf Bottom:}
 The ratio between the SHM
  and the \SHMpp~cross sections. The shading indicates
  the ratio for different values of $\eta$ ($\eta=0.2$ corresponds to
  the ratio for the top panel).
  The black dotted line indicates the difference that arises solely from the
  different values of $\rho_0$ in the SHM and SHM$^{++}$; deviations from this line
  arise from the different velocity distributions.
 } \label{fig:limits}
 \end{figure}

The results of direct detection experiments are usually summarised in
terms of exclusion limits on the SI DM-proton scattering cross section as a function of DM mass.
In Fig.~\ref{fig:limits}, we illustrate the effects of moving from the SHM to the \SHMpp~for three
hypothetical experiments using a xenon (green), germanium (purple) and a He:SF$_6$ (red) target material.
In the upper panel, the dashed lines show the limits for the SHM with parameters in the upper half of Table~\ref{tab:astrobenchmarks}, while the
solid lines show the limits for the SHM$^{++}$ with our new recommended values for the astrophysical
parameters given in the lower half of Table~\ref{tab:astrobenchmarks}.
 The limits are calculated as median discovery
limits, where we use the profile likelihood ratio test under
the Asimov approximation to calculate the cross sections discoverable at~3$\sigma$ 
(see Ref.~\cite{Cowan:2010js} for more details). WIMP 90\% CL exclusion limits
will follow the same behaviour as the discovery limits shown in Fig.~\ref{fig:limits}.

The green limits correspond to a toy version of a liquid xenon experiment like DARWIN~\cite{Aalbers:2016jon}
with a $\sim$200 ton-year exposure.
As a proxy, we have used the background rate
and efficiency curve reported for LZ~\cite{Akerib:2018lyp}. 
The low threshold germanium result (purple limits) is a toy version of the SuperCDMS~\cite{Agnese:2017jvy} 
or EDELWEISS~\cite{Arnaud:2017usi} experiments, where we assume a simple error function parameterisation
for the efficiency curve, which falls sharply towards a threshold at 0.2 keV.
The He:SF$_6$ target (red limits) is a toy version of the 1000m$^3$ CYGNUS
directional detector using a helium and SF$_6$ gas mixture 
(discussed in more detail in Sec.~\ref{sec:directional}). We have also
included realistic estimates of the detector resolutions in our results.

The upper gray shaded regions in Fig.~\ref{fig:limits} show the
existing exclusion limits on the SI WIMP-proton cross section (calculated
assuming the SHM with the parameters in the upper half of Table~\ref{tab:astrobenchmarks}).
 This is an interpolation of the limits of (from low to high masses)
CRESST~\cite{Angloher:2015ewa}, DarkSide-50~\cite{Agnes:2018oej},
LUX~\cite{Akerib:2016vxi}, PandaX~\cite{Tan:2016zwf} and
XENON1T~\cite{Aprile:2018dbl}. 
The lower blue region shows the `neutrino floor' region for a xenon target.
The neutrino floor
delimits cross sections where the neutrino background saturates the DM
signal, so is therefore dependent upon the shape of the signal model
that is assumed~\cite{OHare:2016pjy}. We calculate the floor in the same 
manner as described in Refs.~\cite{Billard:2013gfa,Ruppin:2014bra,OHare:2016pjy}

Fig.~\ref{fig:limits} shows a noticeable shift between the SHM and
\SHMpp~limits. This is mostly due to the different values of $\rho_0$,
which can be most clearly seen from examining the ratio between the
limits shown in the lower panel. The black dotted line in the lower
panel indicates the ratio 0.55/0.3, the ratio of the different
$\rho_0$ values.  It is only as the limits approach the lowest DM mass
to which each experiment is sensitive that the ratio of cross sections
deviate significantly from the black dotted line.  The small impact on
the \emph{shape} of the exclusion limits can be understood as follows.
Contrasting the SHM and \SHMpp~signals, there are two competing
effects which act to push the limits in opposite directions.
Increasing $v_0$ strengthens the cross section
limits because it increases the number of recoil events above
the finite energy threshold. However, the Sausage reverses this effect
since, as we saw in Fig.~\ref{fig:eventrates}, the Sausage component
decreases the maximum recoil energy so there are fewer events above
the finite energy threshold.

The neutrino floor has a more complicated relationship with the
velocity distribution and the WIMP mass. The cross section of the
floor depends upon how much the neutrino background overlaps with a
given DM signal. The neutrino source that overlaps most with a DM
signal depends on $m_\chi$.  This leads to the non-trivial dependence
of the neutrino floor on the Sausage fraction $\eta$ shown in the
lower panel.

Altogether, our refinement of the SHM ultimately leads to only slight
changes to the cross section limits which, for the most part, are
simple to understand.  This can be considered a positive aspect of our
new model, since while it includes refinements accounting for the most
recent data, it simultaneously allows existing limits on DM particle
cross sections to be used with confidence.  The most notable
difference in the limits arises from the larger value of $\rho_0$,
which can be implemented trivially as an overall scaling.  In the
event of the positive detection of a DM signal, which would lead to
closed contours in the mass--cross section plane, the use of the wrong
model could lead to an incorrect bias in the measurement of both the
DM mass and cross section.\footnote{Mitigating strategies are
  possible~\cite{Peter:2011eu,Kavanagh:2013wba,Kavanagh:2016xfi} but
  they require a large number of signal events to be effective.}  Our
refinements to the halo model will be even more important to consider
in the context of a discovery so that any bias is minimised.

\subsection{Directional signals}\label{sec:directional}
\begin{figure*}[t]
\centering \includegraphics[width=\textwidth,trim={1cm 0cm 2cm 0cm},clip]{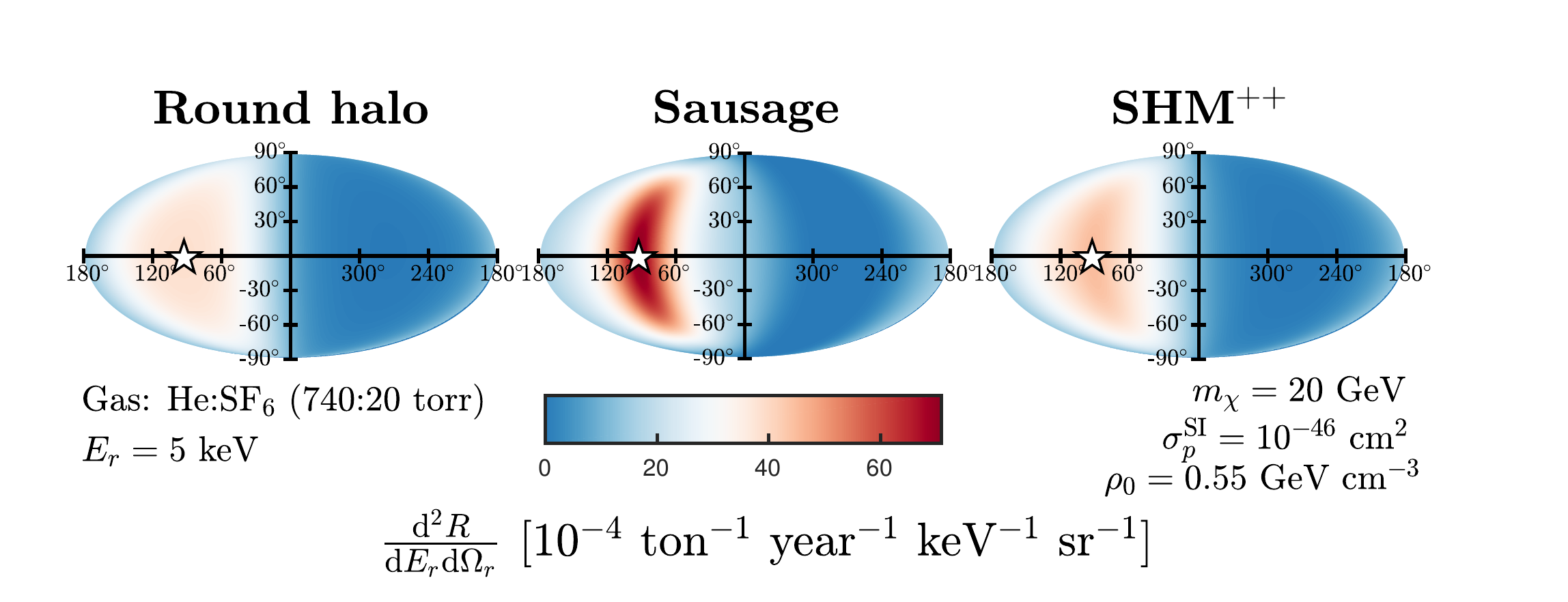}
\caption{Mollweide projection in galactic coordinates of the value of
  the double differential angular recoil rate as a function of the
  inverse of the recoil direction $-\qhat$ at a fixed recoil energy of
  5 keV. We assume a 20 GeV DM mass and sum the rates from both He and SF$_6$. The panels from left to right show the
  distributions for distribution of the round halo component, the
  Sausage, and the combined \SHMpp~respectively.
  The Sausage component gives rise to a distinctive pattern compared to the round halo.
  We indicate
  the direction of $\mathbf{v}_{\mathrm{E}}$ with a white star.
}\label{fig:directional}
\end{figure*}
\begin{figure}[t]
\includegraphics[width=0.49\textwidth]{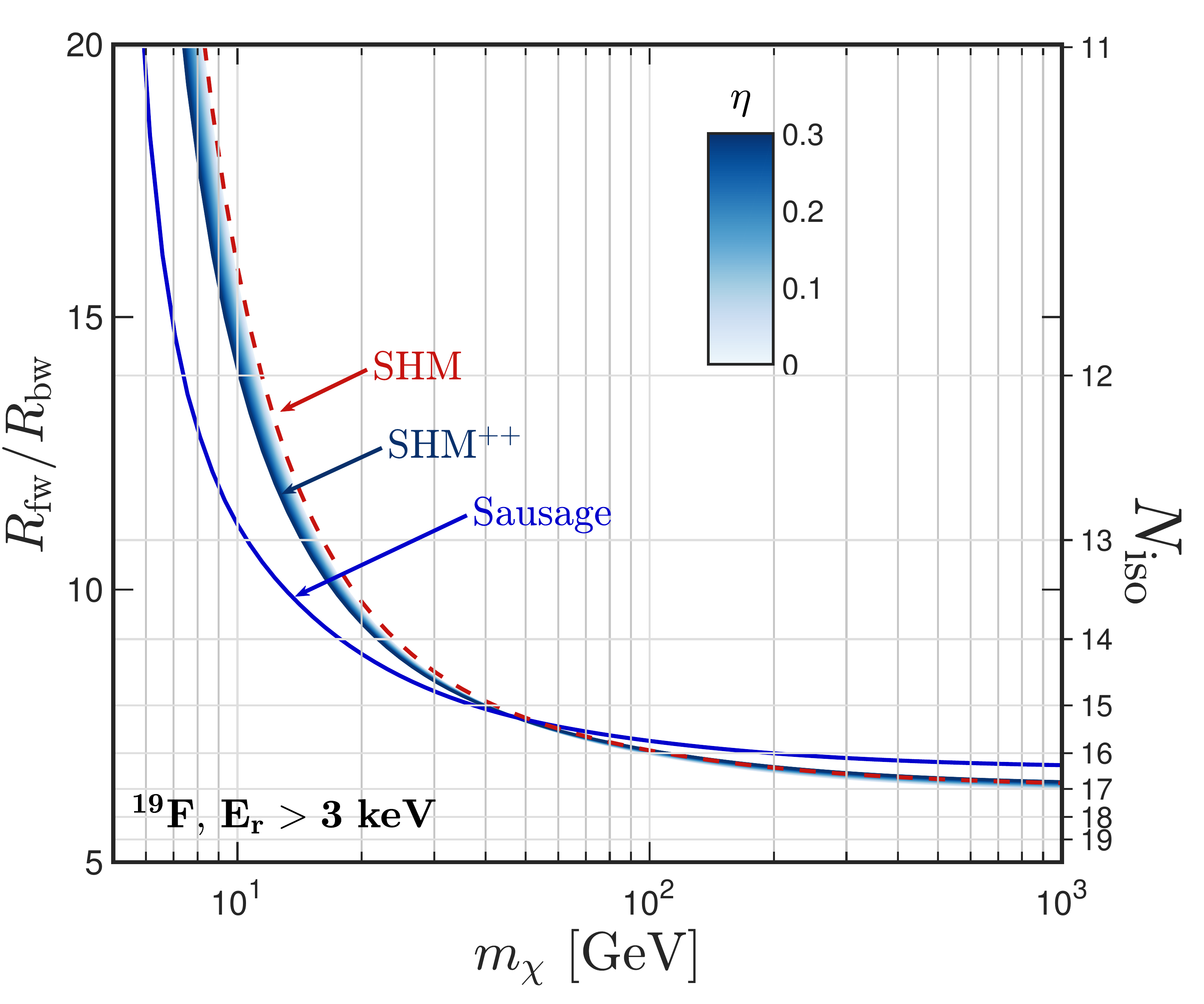}
\caption{Anisotropy of the WIMP directional signal as a function of
  WIMP mass. We quantify this anisotropy as the ratio of the total event
  rate in the forward hemisphere $R_{\rm fw}$ (pointing towards
  $-\mathbf{v}_{\mathrm{E}}$) relative to the rate in the backward
  hemisphere $R_{\rm bw}$ (pointing towards
  $\mathbf{v}_{\mathrm{E}}$). The blue region corresponds to the 
  \SHMpp~with $\eta = 0$ to 30\%. The red dashed line shows the
  anisotropy for the SHM. The anisotropy of the Sausage alone is shown as the
  blue line. Here we integrate above a recoil energy threshold of 3 keV. On the right hand 
  axis we indicate the approximate number of events to detect the anisotropy at the same position on the 
  left hand axis.}\label{fig:anisotropy}
\end{figure}
The main difference between the kinematic structures of the round halo
and the Sausage are at the level of the full three-dimensional velocity
distribution. Much of this structure is integrated away when computing the
speed distribution and the integrals that depend on it. To appreciate the full impact of the Sausage component, we
should consider experiments which are not only sensitive to the speed of
incoming DM particles but also their direction. Such directional
detectors are well motivated on theoretical grounds (see
Ref.~\cite{Mayet:2016zxu} for a review) because the signal from an
isotropic DM halo gives a distribution of recoil angles that aligns
with Galactic rotation~\cite{Spergel:1987kx}, thus clearly
distinguishing it from any background~\cite{Grothaus:2014hja,
  O'Hare:2015mda,OHare:2017rag}. Directional detectors preserve much
more kinematic information about the full velocity
distribution~\cite{Morgan:2004ys,Billard:2009mf,Lee:2012pf,O'Hare:2014oxa,Mayet:2016zxu,Kavanagh:2016xfi}.

For directionally sensitive detectors, the double differential event
rate as a function of recoil energy, recoil direction and time is
proportional to an analogous halo integral, called the Radon
transform~\cite{Gondolo:2002np,Radon},
\begin{equation}
 \hat{f}(\vmin,\qhat,t) = \int \delta\left(\textbf{v} \cdot \qhat - \vmin\right) f(\textbf{v}+\textbf{v}_{\mathrm{E}}(t))\, \textrm{d}^3 \textbf{v}\, ,
\end{equation}
where~$\qhat$ is the direction of the recoiling nucleus.
This enters into an analogous formula to Eq.~\eqref{eq:eventrate} for
the double differential recoil rate with energy and angle,
$\textrm{d}^2R/\textrm{d}E_r \textrm{d}\Omega_r$.

Directional detectors are challenging to build, as
Refs.~\cite{Ahlen:2009ev,Battat:2016xaw} discuss. Many ideas have been proposed to develop detector technologies with angular recoil
sensitivity, including nuclear
emulsions~\cite{Nygren:2013nda,Li:2015zga} and columnar
recombination~\cite{Aleksandrov:2016fyr,Agafonova:2017ajg} for nuclear
recoils, as well as several novel engineered materials for electron
recoils~\cite{Griffin:2018bjn,Hochberg:2016ntt,Hochberg:2017wce}. Experimentally
the most developed technique for is to use gaseous time projection
chambers (see
Refs.~\cite{Daw:2011wq,Battat:2014van,Battat:2016xaw,Monroe:2011er,Leyton:2016nit,Santos:2011kf,Riffard:2013psa,Nakamura:2015iza}).
A large-scale gaseous time projection chamber called CYGNUS has been
proposed and a feasibility study is currently underway~\cite{CYGNUS}.
Two gases are under investigation for CYGNUS: SF$_6$ at 20 torr and
$^4$He at 740 torr.  Both have a total mass of 0.16 tons for a
1000~m$^3$ experiment at room temperature.  Based on these targets we
show $\textrm{d}^2R/\textrm{d}E_r \textrm{d}\Omega_r$, for the round
halo (left), Sausage (middle) and the \SHMpp~(right) in
Fig.~\ref{fig:directional}.  Fixing the recoil energy to $E_r=5$ keV,
we display the full-sky map of recoil angles, which clearly show a
distinctive pattern for the Sausage component when compared with the
round halo. Furthermore this effect is preserved even when the Sausage
is a sub-dominant contribution to the full model.


In Fig.~\ref{fig:anisotropy}, we show the ratio of $R_{\rm fw}$ to $R_{\rm bw}$.
$R_{\rm fw}$ is the total rate for scattering with fluorine above $E_r = 3$~keV
in the hemisphere centred around $-\mathbf{v}_{\mathrm{E}}$, while
$R_{\rm bw}$ is the total rate in the opposite hemisphere.
This ratio therefore gives a measure of the anisotropy of the WIMP
directional signal. Fig.~\ref{fig:anisotropy} shows that the Sausage
component decreases the anisotropy of the WIMP
directional signal, albeit by a modest amount.

%
%

We can also express the same behaviour as a function of $N_{\rm iso}$, 
which is an approximate lower limit to the number of events
 required to detect the dipole anisotropy at 3$\sigma$.\footnote{See Refs.~\cite{Green:2010zm,Billard:2009mf,Mayet:2016zxu} for more sophisticated tests of isotropy.} 
 To detect the anisotropy, we require that the contrast in event numbers in the forward/backward hemisphere 
 ($N_{\rm fw} - N_{\rm bw}$) is greater than the typical 3$\sigma$ random deviation expected under isotropy, 3$\sqrt{N_{\rm fw}+N_{\rm bw}}$. 
 Expressed in terms of event rates gives the formula,
\begin{equation}
N_{\rm iso} \approx \left( 3 \, \frac{R_{\rm fw} + R_{\rm bw}}{R_{\rm fw} - R_{\rm bw}} \right)^2 \, .
\end{equation}
As higher energy recoils typically have smaller
scattering angles, more of the anisotropy of the DM flux is preserved
in the tail of $\textrm{d}R/\textrm{d}E_r$. Hence the anisotropy
increases toward the lowest masses displayed in Fig.~\ref{fig:anisotropy}, where only
 the tail of the recoil energy distribution is above threshold. 
 
 The Sausage component (the blue line in Fig.~\ref{fig:anisotropy}) is considerably
less anisotropic than the round halo. This is because the population
of DM in the Sausage is hotter in the radial direction, meaning a
greater number of recoils scatter away from $\mathbf{v}_{\mathrm{E}}$
above a given energy threshold. This effect is exaggerated at low
masses when the only observable particles from the Sausage are those with the most strongly radial orbits. 
At high masses, when the observable part of the recoil energy spectrum samples a much larger portion of the
velocity distribution, the SHM and \SHMpp~nearly converge. When
looking at the full Radon transform down to much lower~$\vmin$,
the Sausage signal is only slightly more anisotropic than the round halo. This is for two reasons.
Firstly, the increased hotness in the radial direction of the triaxial Gaussian is compensated
by an increased coldness in the tangential direction (aligning with $\mathbf{v}_{\rm E}$). Secondly, much of the 
anisotropy of the signal becomes washed out in the stochastic process of elastic scattering.

Nevertheless, the Sausage is a noticeably different class of feature in the angular distribution of recoils. 
This means that, in the event of a detection, a directional experiment would have a better chance of
distinguishing between the Sausage-less model of the halo and the \SHMpp~compared to an experiment
with no directional information. We anticipate that the Sausage will also have an impact on higher order directional features~\cite{Bozorgnia:2011vc,Bozorgnia:2012eg}, the time integrated signal~\cite{OHare:2017rag}, and angular signatures of operators with transverse velocity dependence~\cite{Kavanagh:2015jma,Catena:2015vpa}, but for brevity we leave these to future studies.

\subsection{Axion haloscopes}\label{sec:axions}
 \begin{figure}[t]
\centering
\includegraphics[width=0.49\textwidth]{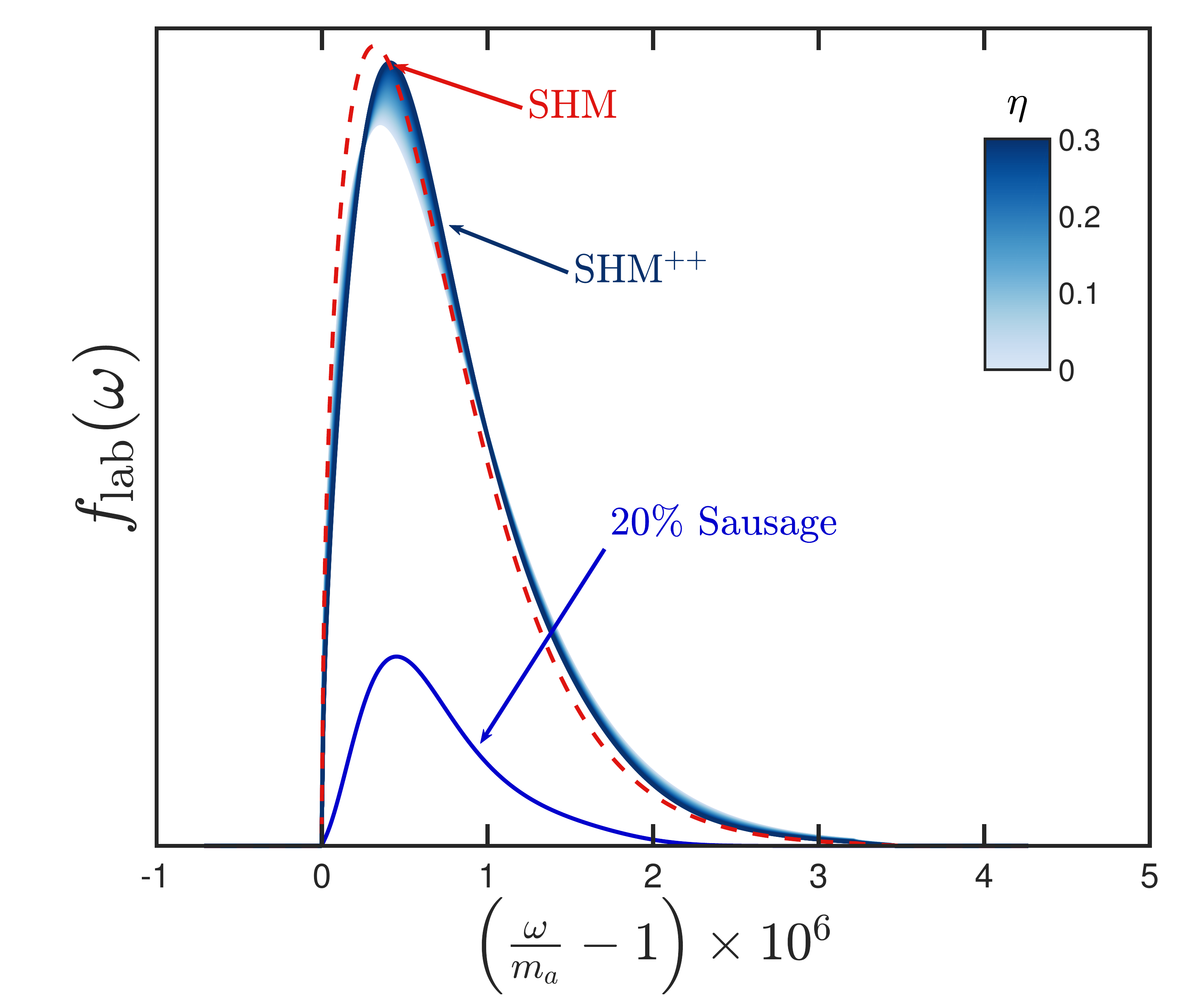}
\caption{Spectral lineshape observable in an axion haloscope. We show
  only the shape of the signal distribution as a function of frequency
  $\omega$, scaled by the axion mass $m_a$. As in previous figures the
  red dashed line shows the SHM, whereas the blue region shows the
  new signal model from the \SHMpp, shaded to indicate the range of
  values of $\eta$. We also isolate a 20\% contribution from the
  Sausage, shown as a blue line.}\label{fig:axionspectrum}
 \end{figure}

The detection of axions is different from WIMPs and requires a different procedure to demonstrate the effect of the new halo model. To detect axions, the standard approach is to attempt to convert them into photons inside the magnetic field of some instrument.
In the event of a detection, the electromagnetic response from axion-photon conversion can be measured in such a device as a function of
frequency. The frequency of the electromagnetic signal is given by
$\omega = m_a(1+v^2/2)$, so the spectral distribution of photons
measured over many coherence times of the axion field oscillations will approach
the astrophysical distribution of speeds on Earth, $f_{\rm{lab}}(v)$ (cf.~Fig.~\ref{fig:fv}).
To identify the frequency of the axion mass, $m_a$, the experiment may either enforce some
resonance or constructive interference condition for a signal
oscillating at $\omega = m_a$
(as in e.g.~ADMX~\cite{Asztalos:2009yp,Du:2018uak},
MADMAX~\cite{TheMADMAXWorkingGroup:2016hpc,MADMAXinterestGroup:2017bgn,Millar:2016cjp},
HAYSTAC~\cite{Brubaker:2016ktl,Rapidis:2017ytq,Brubaker:2017rna,Zhong:2017fot,Brubaker:2018ebj},
CULTASK~\cite{Chung:2016ysi,Lee:2017mff,Chung:2017ibl},
Orpheus~\cite{Rybka:2014cya},
ORGAN~\cite{McAllister:2017lkb,McAllister:2017ern},
KLASH~\cite{Alesini:2017ifp} and RADES~\cite{Melcon:2018dba}), or be
sensitive to a wide bandwidth of frequencies simultaneously
(e.g.\ ABRACADABRA~\cite{Kahn:2016aff,Foster:2017hbq,Henning:2018ogd},
BEAST~\cite{McAllister:2018ndu} and
DM-Radio~\cite{Silva-Feaver:2016qhh}). The axion signal lineshape has
a quality factor of around $10^6$ so even in the best resonant devices,
the full axion signal will be measured at once. This means that in
both resonant and broadband configurations, the sensitivity to axions is
 dependent upon how prominently the signal can show up over a
noise floor. For a recent review of experiments searching for axions
see Ref.~\cite{Irastorza:2018dyq}.

The axion spectral density is proportional to the speed distribution,
up to a change of variables between frequency and speed (see
e.g.\ Refs.~\cite{OHare:2017yze,Foster:2017hbq})
\begin{equation}
\frac{\textrm{d}P}{\textrm{d}\omega} = \pi \mathcal{H}(\omega) \,g^2_{a\gamma} \,\rho_0 \, f_{\rm{lab}}(\omega) \, ,
\end{equation}
where $\mathcal{H}(\omega)$ encodes experimental dependent
factors and $g_{a\gamma}$ is the axion-photon coupling on which the
experiment will set a limit. The shape of the axion signal is dominated by the
term 
\begin{equation}
f_{\rm{lab}}(\omega) =\frac{\textrm{d}v}{\textrm{d}\omega}\, f_{\rm{lab}}(v)\, ,
\end{equation}
since the frequency dependence of $\mathcal{H}(\omega)$ in any realistic experiment will be effectively constant
over the small range of frequencies covered by galactic speeds.
We show $f_{\rm{lab}}(\omega)$ as a function of frequency in
Fig.~\ref{fig:axionspectrum}. This distribution is similar to $f_{\rm{lab}}(v)$, 
which was presented in Fig.~\ref{fig:fv}, 
but is now a function of the observable quantity in an axion experiment. 

The statistical methodology of a generic
axion DM experiment consists of the spectral analysis of a
series of electromagnetic time-stream samples. The stacking of the
Fourier transforms of this time-stream data in most cases leads to
Gaussian noise suppressed by the duration of the experiment, as well
as (ideally) an enhanced axion signal on top. Hence a
likelihood function for such data can be written in terms of a
$\chi^2$ sum over frequency bins which ultimately can be approximated
in terms of the integral over the power spectrum squared.  Since the
power is proportional to $g^2_{a\gamma}$, the minimum discoverable
value scales with the shape of the signal as~\cite{Foster:2017hbq},
 \begin{equation}
g_{a\gamma} \propto \sqrt{\frac{1}{\rho_0}} \left(\int_{m_a}^{\infty} \mathrm{d} \omega\, f_{\rm{lab}}(\omega)^2 \right)^{-1/4} \, .
 \end{equation}
This formula encodes the fact that signals that are sharper in frequency are more prominent over white noise and hence easier to detect. 
However, the dependence on the width of $f_{\rm{lab}}(\omega)$ and therefore the width of $f_{\rm{lab}}(v)$
only enters weakly, as an integral raised to the $-1/4$ power, so although the SHM$^{++}$ distribution is colder, the overall effect is small.
 Additionally, the Sausage component is not especially localized at a
given frequency so again, its impact is small.
For demonstration, a hypothetical experiment
which used an~$\eta = 1$ signal model would set limits on~$g_{a\gamma}$ 
only around 6\% stronger that the
same experiment using $\eta = 0$.

For the parameters in Table~\ref{tab:astrobenchmarks} -- modulo the value $\rho_0 =
0.45 \GeVcm$ instead of $\rho_0 =
0.3 \GeVcm$ in the SHM to reflect the preference of haloscope collaborations --
constraints on $g_{a\gamma}$ assuming the \SHMpp~relative to the the SHM are around 8\% \emph{stronger}. As in the case of WIMPs , there are several competing effects. 
The increase in $v_0$ acts to broaden the signal line-width making constraints weaker. 
However the inclusion of the Sausage component, which is a slightly sharper signal, balances against this. 
Based on the difference in shapes of $f_{\rm{lab}}(v)$ alone, constraints when using the \SHMpp~would be about 2\% \emph{weaker}. 
The final balancing act comes from the new increased value of $\rho_0$. This ultimately has the greatest impact and pushes the \SHMpp~constraint to be stronger than the SHM. 

The sensitivity of axion haloscopes to astrophysics is essentially only controlled by the width of the speed distribution 
(rather than moments above some cutoff as is the case for WIMPs). Hence it is not surprising that the refinements that 
we have made have little impact on limits on the axion-photon coupling. As has been discussed in the past, 
the only changes that can bring significant changes to the axion signal are cold substructures like streams, 
which present highly localised peaks in frequency~\cite{OHare:2017yze,Foster:2017hbq,Knirck:2018knd,OHare:2018trr}. 
One exception may be directional axion experiments possessing sensitivity to the full velocity distribution via prominent diurnal modulations~\cite{Irastorza:2012jq,Knirck:2018knd}. 
These will be altered significantly by the Sausage component, however such experiments remain hypothetical.

\section{Conclusions}\label{sec:conc}

The data from the \emph{Gaia} satellite~\cite{GaiaDR2} has driven many
changes in our picture of the Milky Way Galaxy. Firstly, more
prosaically, it has enabled the uncertainties in many Galactic
parameters to be substantially reduced. The halo shape
(Sec.~\ref{sec:shape}), circular speed (Sec.~\ref{sec:v0}) and the
escape speed (Sec.~\ref{sec:vesc}) are now much more securely pinned
down than before. Only the local DM density (Sec.~\ref{sec:rho0})
remains obdurately uncertain, though analyses in the near future of
\emph{Gaia} Data Release~2 should improve constraints on its value.

 Secondly and more spectacularly, it has provided unambiguous evidence
 of an ancient head-on collision with a massive
 ($10^{10}-10^{11}\Msol$) satellite galaxy
 ~\cite{Kr18,My18,Be18,Ma18}, reinforcing earlier suggestions that the
 local halo is bimodal~\cite{Ca07}. The stellar debris from this event
 encompasses our location, with many of the stars moving on strongly
 radial orbits. In addition to stars, the satellite galaxy will have
 disgorged huge amounts of DM, having a radical effect on the velocity
 distribution.

The standard halo model (SHM) has provided trusty service in
astroparticle physics as a representation of the Milky Way halo that
is both simple and realistic.  We have put forward here its natural
successor, the \SHMpp, in which the Galactic parameters are updated in
the light of the advances from \emph{Gaia} Data Release~2 and the dark
halo's bimodal structure is explicitly acknowledged. Each of the two
components can be modelled as Gaussian, though the Sausage is strongly
radially anisotropic. The combined velocity distribution is of course
not Gaussian, as illustrated in Fig.~\ref{fig:fv}, but nevertheless
it remains easy to use and manipulate. Compared to the SHM, there are
two additional parameters, namely the fraction of DM~$\eta$ and the
velocity anisotropy~$\beta$ of the DM in the Sausage. These parameters
can be constrained from astrophysical arguments to $10\% \lesssim \eta
\lesssim 30 \% $ and $\beta \approx 0.9\pm 0.05$.  A succinct comparison
between the SHM and \SHMpp~is given in
Table~\ref{tab:astrobenchmarks}. We have given recommended central
values and 1$\sigma$ uncertainties. For measured parameters, these are
motivated by existing statistical uncertainties, but for the
parameters of the dark Sausage we have used theoretical arguments.

We have computed the effects of the \SHMpp~on a range of DM
experiments, comparing our results to the benchmark SHM. The addition
of the radially anisotropic Sausage makes the velocity distribution
colder. However, this is compensated by the increase in the local
circular speed from 220 to 235 kms$^{-1}$, making the velocity
distribution hotter. This explains why the change in the rate of
nuclear recoils in direct detection experiments, for example, is
modest. We demonstrated these effects on the halo integrals $g(\vmin)$
and $h(\vmin)$ (Fig.~\ref{fig:gvmin}) which control the halo
dependence of direct detection signals, as well as on the observable
distribution of recoil energies (Fig.~\ref{fig:eventrates}).  In the
context of particle physics measurements for WIMPs, the projected
exclusion limits, as well as the neutrino floor, are very similar for
both SHM and \SHMpp~(Fig.~\ref{fig:limits}). In fact, the dominant
change here is the factor of $\sim1.8$ increase in sensitivity due to
the updated value of $\rho_0$ from 0.3 to 0.55$\GeVcm$.

Examination of other signals shows a similar pattern. Like the round
halo, the Sausage is centred at the origin in velocity space, so the
relative velocity between the DM and Earth rest frames oscillates with
the same phase, and hence the annual modulation signal is left largely
unchanged (Fig.~\ref{fig:annualmodulation}).  The instances in which
it is most important to use the bimodal \SHMpp~are experiments that
are explicitly sensitive to the three-dimensions of the velocity
distribution. We studied this type of signal for a future directional
WIMP search like CYGNUS~\cite{CYGNUS}.  The Sausage component leaves a
distinctive recoil angle distribution in directional WIMP searches
(Fig.~\ref{fig:directional}).  The \SHMpp~is less anisotropic than the
SHM (Fig.~\ref{fig:anisotropy}).  This may raise the concern that the
\SHMpp~might weaken prospects for the directional discovery of DM but
the change in the number of events for a detection is only marginally
increased.

As Fig.~\ref{fig:limits} demonstrates, DM-nucleon cross section limits
calculated assuming the older SHM are similar to the limits from
\SHMpp~(after the rescaling from the different values of~$\rho_0$ have
been taken into account).  This means that older exclusion limits are
still reasonably accurate.  We find a similar result also holds for
axion haloscopes: constraints on the axion-photon coupling would be
only marginally stronger with the \SHMpp.  Crucially, since the width
of the axion signal (set by the width of the speed distribution) is
similar in the SHM and~\SHMpp~(Fig.~\ref{fig:axionspectrum}), there
cannot be major changes to signals in axion haloscopes.  Our overall
recommendation is that the~\SHMpp~should be adopted in future direct
detection searches, since it retains much of the simplicity of the SHM
while more accurately capturing the known properties of the Milky Way
halo. However, for experiments without directional sensitivity,
acceptable results may be obtained simply by updating the SHM with the
new values for $\rho_0$, $v_0$, and $\vesc$.

Finally, while this new model represents a well-motivated elaboration
of the SHM for the purposes of direct detection analyses, it may not
be the final word on the local structure of the DM
distribution. Importantly there likely will be substructure in the
velocity distribution~\cite{Vogelsberger:2010gd,Lisanti:2010qx} with
consequences for direct detection
experiments~\cite{Lee:2012pf,O'Hare:2014oxa,Savage:2006qr,Foster:2017hbq,OHare:2017yze,Kavanagh:2016xfi}. In
fact, the S1 tidal stream was recently spotted in \emph{Gaia}
data~\cite{Myeong:2017skt} and clearly intersects the Solar
position~\cite{Myeong:Preprint}.  The potentially observable signals
in the next generation of DM experiments have been
investigated~\cite{OHare:2018trr} and its effects may be more
significant than the Sausage. This is especially true for directional
WIMP and axion experiments, as the S1 stellar stream is strongly
retrograde and its velocity signature is unlike the smooth halo.  We
have not included the S1 stream in the \SHMpp~because we cannot
currently constrain its contribution to $\rho_0$.  However, the next
refinement may be to incorporate the S1 stream, though this must await
a more complete understanding of the DM component of the stream (as
well as the relationship between DM and stellar populations in
general).

\acknowledgments We thank Henrique Araujo, Nassim Bozorgnia, Anne
Green and Mariangela Lisanti for their comments on an early draft of
this paper. NWE thanks Vasily Belokurov and Nicola Amorisco for many
interesting Sausage discussions. CAJO is supported by the grant
FPA2015-65745-P from the Spanish MINECO and European FEDER.  CM is
supported by the Science and Technology Facilities Council (STFC)
Grant ST/N004663/1. This work was partly performed at the Aspen Center
for Physics, which is supported by National Science Foundation grant
PHY-1607611 and by a grant from the Simons Foundation.

\bibliographystyle{apsrev4-1}
\bibliography{New_SHM}

\end{document}